\documentclass[aip,reprint,pop]{revtex4-1}

\usepackage{graphicx}
\usepackage{subfigure}
\usepackage{epstopdf}
\usepackage{amsmath}
\usepackage{nicefrac}
\usepackage{dcolumn}
\usepackage{bm}
\usepackage{mathrsfs}
\usepackage[colorlinks=true, linkcolor=blue]{hyperref}

\DeclareMathOperator{\sech}{sech}

\newcommand{\deltaMa}{\mathop{}\!\delta\!M_A}

\begin{document}

\title{Cumulative displacement induced by a magnetosonic soliton bouncing in a bounded plasma slab}
\author{Renaud Gueroult}
\affiliation{LAPLACE, Universit\'{e} de Toulouse, CNRS, 31062 Toulouse, France}
\author{Amnon Fructhman}
\affiliation{Faculty of Sciences, H.I.T.-Holon Institute of Technology, Holon 58102, Israel}
\author{Nathaniel J. Fisch}
\affiliation{Princeton Plasma Physics Laboratory, Princeton University, Princeton, NJ 08543 USA}

\begin{abstract}
The passage of a magnetosonic (MS) soliton in a cold plasma leads to the displacement of charged particles in the direction of a compressive pulse and in the opposite direction of a rarefaction pulse. In the overdense plasma limit, the displacement induced by a weakly nonlinear MS soliton is derived analytically. This result is then used to derive an asymptotic expansion for the displacement resulting from the bouncing motion of a MS soliton reflected back and forth in a vacuum-bounded cold plasma slab. Particles' displacement after the pulse energy has been lost to the vacuum region is shown to scale as the ratio of light speed to Alfv{\'e}n velocity. Results for the displacement after a few MS soliton reflections are corroborated by particle-in-cell simulations.
\end{abstract}

\date{\today}
\maketitle

\section{Introduction}

A singular subset of nonlinear waves are waves for which dispersion balances the wave-steepening effects that arise from nonlinearity. In weakly dispersive media, the propagation of these nonlinear waves can be described by the Korteweg-de-Vries (KdV) equation~\cite{Korteweg1895}. The KdV equation can have two kinds of stationary solutions: periodic cnoidal waves~\cite{Korteweg1895,Wiegel1960} and solitary localised waves, or solitons~\cite{Zabusky1965}. Solitons are remarkable objects in that they preserve their shape and speed after collision, behaving in some ways like particles~\cite{Belashov2006}. 

KdV equations have been derived both for ion-acoustic wave~\cite{Washimi1966} and for magnetosonic (MS) wave~\cite{Gardner1965} in homogeneous unmagnetized and magnetized plasmas, respectively. For plasmas featuring multiple ion species, both the ion-acoustic wave and the MS wave splits into a fast and a slow mode~\cite{Mikhailovskii1985}, and each of these four modes can in turn be described by a separate KdV equation~\cite{Tran1974a,Ohsawa1985c,Toida1994}. Solitary waves matching the properties of soliton solution to the KdV equation for ion-acoustic waves have been produced in laboratory experiments~\cite{Ikezi1970} while solitary waves matching the property of soliton solution to the KdV equation for slow MS waves have been observed in space plasmas~\cite{Stasiewicz2003}.

The realization that MS solitons can describe the initial state of the formation of subcritical perpendicular shocks~\cite{Sagdeev1966,Tidman1971,Biskamp1973,Balogh2013,Ohsawa2014,Gueroult2017} motivated the study of the structure of nonlinear MS waves~\cite{Davis1958,Adlam1958,Adlam1960,Banos1960,Hain1960}. Following these early studies, a particular focus has been on the particle dynamics in large amplitude nonlinear MS waves, both solitary~\cite{Ohsawa1986a,Ohsawa1990,Rau1998,Maruyama1998,Stasiewicz2007,Ohsawa2017a} and periodic~\cite{Lembege1983,Lembege1984,Lembege1986,Lembege1989}, to uncover acceleration mechanisms which could explain the observation of energetic particles in astrophysics~\cite{Drury1983}.

Besides acceleration, another effect of the passage of a MS soliton is to displace particles. Indeed, as noted by Adlam and Allen~\cite{Adlam1958}, ``\emph{the plasma returns to its initial state after the passage of the wave, except that each particle has been displaced in the direction of propagation}''. That the passage of a soliton displaces particles might be of little interest in astrophysical settings, which may be why this effect has received limited attention. On the other hand, the ability to control plasma displacement and, in turn, plasma position is desirable in various laboratory plasma experiments, such as magnetic confinement fusion experiments~\cite{Yuan2013} and non-neutral plasmas~\cite{Canali2011}. One possible control mechanism may lie in the plasma displacement induced by a magnetosonic wave. Compressional Alfv{\'en} waves produced by dedicated magnetic coils have for example been suggested to stabilize plasmas in mirror machines~\cite{Lifshitz2012}. Yet, soliton propagation, and more generally wave propagation, in laboratory plasmas differs from the situation considered in space plasmas in that laboratory plasmas are of finite spatial extension and bounded.


The presence of physical boundaries in laboratory plasmas leads to sheaths where the plasma is inhomogeneous~\cite{Lieberman1994}. Since the KdV equation is only valid for homogeneous plasmas~\cite{Gardner1965,Washimi1966}, wave propagation in these regions cannot be described by a KdV equation, and the stationary soliton solutions are not valid. Yet, for slowly varying media, \emph{i. e.} weak gradients, reductive perturbation technique~\cite{Taniuti1968} can be used to derive a modified KdV (mKdV) equation both for ion-acoustic~\cite{Ko1978} and for MS~\cite{Kakutani1971} waves. Perturbative theory predicts that solitons will no longer be stationary and that an oscillatory tail will form behind the soliton~\cite{Berezin1967,Tappert1971,Karpman1977,Karpman1978,Ko1980}. For stronger gradients, a soliton may be reflected~\cite{Nakata1988,Lonngren1991}. Strong reflection of an ion-acoustic soliton by the sheath formed in front of biased grid electrodes has for example been reported~\cite{Dahiya1978,Nishida1984,Imen1987}. By applying suitable boundary conditions, a soliton can then be forced to bounce back and forth in a laboratory plasma, as it was demonstrated for an ion-acoustic soliton~\cite{Cooney1991}.

In this paper, we investigate how a MS soliton bounces within a magnetized plasma slab bounded by vacuum with the goal of assessing the displacement of particles induced by the soliton's repetitive passages. By considering a 1d plasma slab immersed in a perpendicular background magnetic field, particles are confined without the need for physical boundaries. This allows us to consider the plasma slab homogeneous in first approximation. At the plasma-vacuum boundaries, plasma density drops to zero over a few Debye lengths, and this sharp transition reflects the incident MS soliton~\cite{Nakata1988}.  

The paper is organized as follows. In Sec.~\ref{Sec:SinglePassage}, we derive, to our knowledge for the first time, the displacement induced by the passage of a small amplitude MS soliton in the overdense regime. In Sec.~\ref{Sec:Infinite}, we use this result to derive the displacement produced by an infinite number of bounces. In Sec.~\ref{Sec:PIC}, we validate our analytical findings through particle-in-cell simulations. In Sec.~\ref{Sec:Summary}, the main findings are summarized.

\section{Particle displacement induced by a magnetosonic soliton}
\label{Sec:SinglePassage}

We first consider the plasma displacement induced by a nonlinear magnetosonic (MS) solitary wave propagating along the $x$ direction. Calculations are carried out in the wave frame, with the wave travelling at a velocity $-V_0$ in the negative $x$ direction. 

\paragraph{Longitudinal electric field. -- } Introducing $B_0$ and $n_0$ the unperturbed magnetic field and density [$B_z(-\infty) = B_0, n(-\infty) = n_0$], and following Ref.~\cite{Rau1998}, the normalized longitudinal electric field $E = E_x/(B_0 c)$ is related to the normalized magnetic field $B = B_z/B_0$ by the bi-quadaratic equation
\begin{equation}E^4+a_1(B) E^2 + a_0(B) = 0
\label{Eq:E_quadratic}
\end{equation}
where
\begin{subequations}
\begin{multline}
a_1(B) = 2\left[1-B^2+2\beta^2+2{M_A}^2(1+2\eta^2)\vphantom{\frac{1}{2}}\right.\\\left.+2{M_A}^4\frac{\eta^2(1+\eta^2)}{(\beta B)^2}\right],
\end{multline}
\begin{equation}
a_0(B) = (B^2-1)^2-4{M_A}^2(B-1)^2.
\end{equation}
\end{subequations}
Here, $\beta = V_0/c$ is the normalized wave speed, $M_A = V_0/V_A$ is the Alfv{\'e}n Mach number with $V_A = B_0/\sqrt{\mu_0 n_0 m_p}$ the Alfv{\'e}n speed, and $\eta^2 = m_e/m_p$ is the electron to ion mass ratio. Eq.~(\ref{Eq:E_quadratic}) has solution for $E$ for $B\leq B_m$, with $B_m = 2 M_A-1$. Reproducing Eq.~($12$) from Ref.~\cite{Rau1998}, the magnetic field $B$ verifies
\begin{widetext}
\begin{multline}
\frac{\partial B}{\partial s} = -\frac{\eta B E M_A}{\beta(1-B^2+E^2+2{M_A}^2)(\beta^2B^3+B^2\eta^2{M_A}^2-E^2\eta^2{M_A}^2)}\\ \times\left[2{M_A}^4\eta^2(1+\eta^2)+2{M_A}^2\beta^2B(B+2\eta^2)+B^2\beta^2(1-B^2+2\beta^2+E^2)\right],
\label{Eq:Bexact}
\end{multline}
\end{widetext}
with $s = x \omega_{pe}/c$ the position normalized by the electron skin depth $\lambda_{sd} = c/\omega_{pe}$.
In the limit of a weakly non-linear wave ($B-1\ll1$, \emph{i.~e.} $M_A-1\ll1$) in the over-dense regime ($\omega_{pe}/\omega_{ce}\gg1$, \emph{i. e.} $\eta M_A/\beta\gg1$), the magnetic field can be approximated~\cite{Rau1998} by
\begin{equation}
\bar{B}(s) = 1+2\deltaMa\sech^2\left[s\sqrt{\deltaMa(1+\eta^2)/2}\right],
\label{Eq:barB}
\end{equation}
where $\deltaMa = M_A-1$. One recovers the relation between magnetic field amplitude $B_m$ and Mach number, $B_{m} = 1 +2 \deltaMa$, or $\delta B = 2\deltaMa$, derived from Eq.~(\ref{Eq:E_quadratic}). The magnetic field profile $\bar{B}$ defined in Eq.~(\ref{Eq:barB}) is typical of small amplitude MS compressive solitons~\cite{Rau1998,Ohsawa1986a,Ohsawa1987,Balogh2013,Ohsawa2014}. However, we note that, depending on whether the Alfv{\'e}n speed definition accounts for electron inertia or not, $\sqrt{1+\eta^2}$ is sometimes omitted in the argument of the hyperbolic secant in Eq.~(\ref{Eq:barB}). Consistently with soliton theory, the width of the pulse scales as the inverse of the square root of its amplitude, and the pulse can be described by a single parameter ($\deltaMa$ is used here). Solving Eq.~(\ref{Eq:E_quadratic}) for $E$ with $B=\bar{B}$, and expanding the solution for $\deltaMa \ll 1$ gives
\begin{equation}
\bar{E}(s)=E_0\frac{\sinh(s^{\star})}{\cosh^3(s^{\star})},
\label{Eq:Ebar}
\end{equation}
with 
\begin{subequations}
\begin{equation}
E_0 = (2\deltaMa)^{3/2}\beta/\eta
\label{Eq:Ebar_amp}
\end{equation}
and 
\begin{equation}
s^{\star} = s\left[\deltaMa(1+\eta^2)/2\right]^{1/2}.
\label{Eq:Ebar_width}
\end{equation}
\end{subequations}
Noting that $\textrm{max}[\tanh(u)\sech^2(u)] = 2/(3\sqrt{3})$, the maximum normalized electric field is $2/(3\sqrt{3}){(2\deltaMa)^{3/2}}\beta\eta^{-1}$, which is consistent with the first order term of the asymptotic development given in Eq.~($14$) in Ref.~\cite{Rau1998}, with Eq.~($6$) in Ref.~\cite{Toida1999a} for a single ion species plasma in the low Mach number limit (\emph{i. e.} $\beta \sim V_A /c$), and with Eq.~($67$) in Ref.~\cite{Ohsawa1986a} in the limit of cold plasma and small $\deltaMa$. Integration of Eq.~(\ref{Eq:Ebar}) yields the approximate electric potential
\begin{equation}
\bar{\phi}(s) = \frac{2\deltaMa \beta}{\eta\sqrt{1+\eta^2}}\sech^2(s^{\star}).
\end{equation}

\begin{figure}
\begin{center}
\includegraphics[]{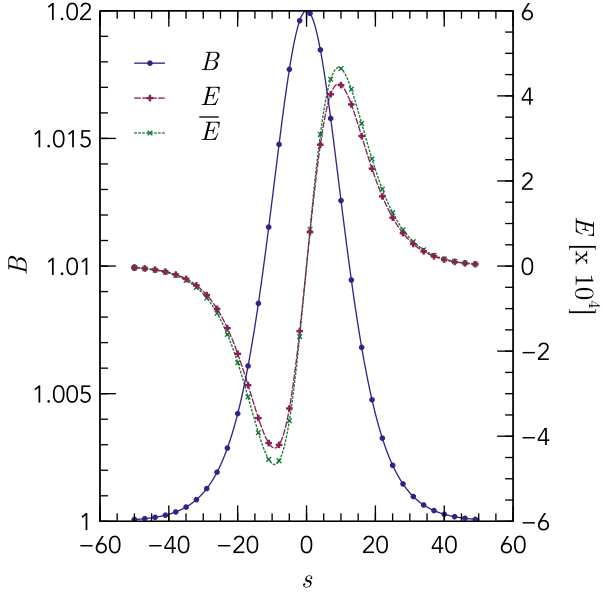}
\caption{Magnetic and electric field profiles as a function of the normalized position $s = x/\lambda_{sd}$ for $\deltaMa = 10^{-2}$, $\eta = 1/\sqrt{1836}$, and $\beta = 10^{-2}$. $E$ is the numerical solution to Eq.~(\ref{Eq:E_quadratic}), whereas $\bar{E}$ is the approximate solution obtained from Eq.~(\ref{Eq:Ebar}).}
\label{Fig:Fields}
\end{center}
\end{figure}

\paragraph{Ion displacement. -- } The velocity of a MS soliton is $M_A V_A$, while its width is $\lambda_{sd}\sqrt{2/\deltaMa}$. The interaction time of a particle with this pulse is therefore
\begin{equation}
\tau_r=\frac{\sqrt{2}\lambda_{sd}}{\sqrt{\deltaMa}M_A V_A}=\frac{\sqrt{2}}{M_A}\sqrt{\frac{\eta^2}{\deltaMa}}\frac{1}{\omega_{ci}},
\label{Eq:PulseInteractionTime}
\end{equation}
with $\omega_{ci}=eB_0/m_p$ the ion cyclotron frequency. Since typically $\eta^2\ll\deltaMa$, $\tau_r\ll\omega_{ci}^{-1}$ and an ion is hence to first order unmagnetized while it interacts with the pulse. The equation of motion for such an ion, initially at rest in the laboratory frame, passing through the pulse defined by Eq.~(\ref{Eq:Ebar}) writes
\begin{equation}
\ddot{s}-\frac{eB_1c}{m_p\lambda_{sd}}E_0\frac{\sinh\left[s\sqrt{\deltaMa(1+\eta^2)/2}\right]}{\cosh^3\left[s\sqrt{\deltaMa(1+\eta^2)/2}\right]} = 0,
\label{Eq:motion}
\end{equation}
with $e$ the elementary charge.
Introducing 
\begin{equation}
\alpha = \frac{2{\deltaMa}^2 \beta}{\eta}\sqrt{1+\eta^2}\frac{eB_1c}{m_p\lambda_{sd}}
\end{equation}
and $\chi = s\sqrt{\deltaMa(1+\eta^2)/2}$ leads, after integration, to
\begin{equation}
\dot{\chi}^2 = -\alpha \sech^2(\chi)+\dot{\chi_0}^2,
\label{Eq:PDE}
\end{equation}
where use has been made of the initial conditions $\dot{\chi}(0) = \dot{\chi_0} = V_0/\lambda^{\star}$ with $\lambda^{\star} = \lambda_{sd}\left[\deltaMa(1+\eta^2)/2\right]^{-1/2}$, and $\chi(0) = \chi_0 = -\infty$. Noting here that 
\begin{align}
\delta & = \alpha/(2\dot{\chi_0}^2) = \frac{2\deltaMa}{\sqrt{1+\eta^2}M_A} \nonumber\\& = \frac{2\deltaMa}{\sqrt{1+\eta^2}}- \frac{2{\deltaMa}^2}{\sqrt{1+\eta^2}} + \mathcal{O}({\deltaMa}^3),
\end{align}
Eq.~(\ref{Eq:PDE}) can be approximated by
\begin{equation}
\dot{\chi} +\frac{\alpha}{2\dot{\chi_0}} \sech^2(\chi) - \dot{\chi_0} = 0.
\label{Eq:PDE_approx}
\end{equation}
Using the variable transform $\zeta = \dot{\chi_0}t-(\chi-\chi_0)$, Eq.~(\ref{Eq:PDE_approx}) writes
\begin{equation}
-\dot{\zeta}+\frac{\alpha}{2\dot{\chi_0}}\sech^2(\chi_0+\dot{\chi_0}t-\zeta) = 0,
\label{Eq:zetadiff}
\end{equation}
which, with the initial condition $\zeta(0) = 0$, can be integrated to give
\begin{multline}
-\zeta + \sqrt{\frac{\delta}{1-\delta}}\arctan\left[\sqrt{\frac{\delta}{1-\delta}}\tanh(\chi_0+\dot{\chi_0}t-\zeta)\right] \\= \sqrt{\frac{\delta}{1-\delta}}\arctan\left[\sqrt{\frac{\delta}{1-\delta}}\tanh(\chi_0)\right].
\label{Eq:zeta}
\end{multline}
The ion displacement along $x$ in the laboratory frame resulting from the passage of the compressive wave is $-\Delta \zeta$, with
\begin{equation}
\Delta\zeta^c = \lim_{t \to \infty} \zeta = 2\sqrt{\frac{\delta}{1-\delta}}\arctan\left[\sqrt{\frac{\delta}{1-\delta}}\right].
\label{Eq:displacement0}
\end{equation}
Expanding for $\delta=2\deltaMa(1+\eta^2)^{-1/2}{M_A}^{-1}\ll1$, Eq.~(\ref{Eq:displacement0}) gives  
\begin{equation}
\Delta \zeta^c = \frac{4}{\sqrt{1+\eta^2}}\deltaMa + \frac{4}{3}\frac{4-3\sqrt{1+\eta^2}}{1+\eta^2}{\deltaMa}^2+\mathcal{O}({\deltaMa}^3). 
\label{Eq:displacement}
\end{equation}

\begin{figure}
\begin{center}
\includegraphics[]{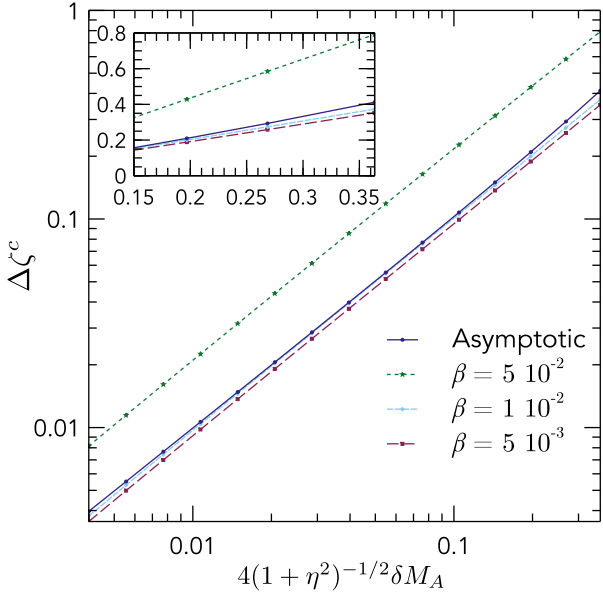}
\caption{Displacement $\Delta \zeta^c$ for various values of $\beta$ as obtained from the Eqs.~(\ref{Eq:E_quadratic}) and (\ref{Eq:Bexact}), and from solving the asymptotic problem described by Eq.~(\ref{Eq:zeta}).}
\label{Fig:Push}
\end{center}
\end{figure}

Comparing the ion displacement after the passage of a single pulse as obtained by solving Eq.~(\ref{Eq:E_quadratic}) and Eq.~(\ref{Eq:Bexact}), and from Eq.~(\ref{Eq:zeta}), indicates, as shown in Fig.~\ref{Fig:Push}, that the asymptotic solution remains within roughly $10\%$ of the exact solution up to $\deltaMa\sim0.05$, granted that $\beta\leq10^{-2}$. This condition on $\beta$ results from the over-dense regime assumption, which can be written as $\eta M_A/\beta\gg1$. By symmetry, the ion displacement in the laboratory frame resulting from the passage of a right propagating compressive pulse is $\Delta \zeta^c$. 

In dimensional units, the ion displacement resulting from the passage of a right propagating compressive pulse is 
\begin{multline}
\Delta x^c = \frac{c\sqrt{B_m-1}}{\omega_{pe}}\left[\frac{4}{1+\eta^2}+\frac{2}{3}\frac{4-3\sqrt{1+\eta^2}}{(1+\eta^2)^{3/2}}(B_m-1)\right.\\\left.+\mathcal{O}\left({(B_m-1)}^{2}\right)\right],
\label{Eq:displacementDimensional}
\end{multline} 
and we write
\begin{equation}
\Delta x_0 = \frac{c}{\omega_{pe}}\frac{4}{1+\eta^2}\sqrt{B_m-1}
\label{Eq:displacementDimensionalZero}
\end{equation}
the first order expansion of $\Delta x^c$.

Although rarefaction pulses of the form
\begin{equation}
\bar{B}_r(s) = 1-2\deltaMa\sech^2\left[s\sqrt{\deltaMa(1+\eta^2)/2}\right]
\label{Eq:barBrar}
\end{equation}
are not solution to the KdV equation for perpendicular magnetosonic wave (see Appendix~\ref{Sec:KdV} and Eq.~(\ref{Eq:KdV})) and therefore do not strictly maintain form while propagating, it is interesting to consider how the ion displacement differs from Eqs.~(\ref{Eq:displacement}, \ref{Eq:displacementDimensional}) in the case of a rarefaction pulse. 
For a rarefaction pulse, the ion motion is in the direction opposed to the pulse propagation. The ion longitudinal displacement then verifies
\begin{equation}
\dot{\zeta}+\frac{\alpha}{2\dot{\chi_0}}\sech^2(\chi_0+\dot{\chi_0}t-\zeta) = 0,
\label{Eq:zetadiff_reverse}
\end{equation}
which has been obtained by reversing the longitudinal electric field in Eq.~(\ref{Eq:zetadiff}). Eq.~(\ref{Eq:zetadiff_reverse}) can be integrated to give
\begin{multline}
-\zeta + \sqrt{\frac{\delta}{1+\delta}}\arctan\left[\sqrt{\frac{\delta}{1+\delta}}\tanh(\chi_0+\dot{\chi_0}t-\zeta)\right] \\= \sqrt{\frac{\delta}{1+\delta}}\arctan\left[\sqrt{\frac{\delta}{1+\delta}}\tanh(\chi_0)\right],
\label{Eq:zeta_reverse}
\end{multline}
which leads to 
\begin{equation}
\Delta\zeta^r = \lim_{t \to \infty} \zeta = 2\sqrt{\frac{\delta}{1+\delta}}\arctan\left[\sqrt{\frac{\delta}{1+\delta}}\right].
\label{Eq:displacement0_reverse}
\end{equation}
Similarly, expanding for $\delta\ll1$, Eq.~(\ref{Eq:displacement0_reverse}) gives  
\begin{equation}
\Delta \zeta^r = \frac{4}{\sqrt{1+\eta^2}}\deltaMa - \frac{4}{3}\frac{8+3\sqrt{1+\eta^2}}{1+\eta^2}{\deltaMa}^2+\mathcal{O}({\deltaMa}^3),
\label{Eq:displacement_reverse}
\end{equation}
or, in dimensional units,
\begin{multline}
\Delta x^r = \frac{c\sqrt{B_m-1}}{\omega_{pe}}\left[\frac{4}{1+\eta^2}-\frac{2}{3}\frac{8+3\sqrt{1+\eta^2}}{(1+\eta^2)^{3/2}}(B_m-1)\right.\\\left.+\mathcal{O}\left({(B_m-1)}^{2}\right)\right].
\label{Eq:displacementDimensional_reverse}
\end{multline}
The effect of the passage of a compressive and a rarefaction soliton is the same to the first order in $\deltaMa$. However, the displacement is enhanced in a compressive pulse ($\Delta x^c \geq \Delta x_0$) since an ion is pushed along the pulse which increases its interaction time with the pulse. The opposite effect is found for a rarefaction pulse, and $\Delta x^r \leq \Delta x_0$.

Note that for the over-dense plasma regime considered here quasi-neutrality holds to second order in $\beta/\eta$  (see, \emph{e.~g}, Refs.~\cite{Adlam1958,Ohsawa2014}). The ion and electron velocity along $x$ is hence the same. As a result, the displacement derived in Eq.~(\ref{Eq:displacement}) and Eq.~(\ref{Eq:displacement_reverse}) not only holds for ions but also for electrons, and those are therefore the plasma displacement for a compression and a rarefaction pulse, respectively.

It is also interesting to note in passing here that quasi-neutrality combined with the soliton definition given by Eq.~(\ref{Eq:barBrar}) is sufficient to recover the equation for the ion motion, Eq.~(\ref{Eq:motion}). Indeed, in the wave frame moving with velocity $-V_0 \hat{x}$, $E_y = V_0 B_0$, and thus the $x$ component of the electron velocity is $v_x = E_y/B_z = V_0/\bar{B}$. Quasi-neutrality implies that $v_x$ is also the $x$ component of the ion velocity. The $y$ component of the Lorentz force on an ion, $e(E_y-v_x\bar{B}B_0)$, is hence zero. It yields that the $y$ component of the ion velocity is zero, which in turn means that the $x$ component of the Laplace force on an ion is zero. As a result, the ion motion along $x$ in the wave frame only depends on $E_x$. In addition, the ion momentum equation along $x$ gives
\begin{equation}
m_p v_x {v_x}' = -\frac{m_p{V_0}^2}{\bar{B}^3} \frac{d \bar{B}}{dx}= e E_x.
\end{equation}
Plugging in the definition of $\bar{B}$ from Eq.~(\ref{Eq:barBrar}), the normalized longitudinal electric field $E = E_x/(B_0c)$ writes
\begin{equation}
E  = 2\frac{\sqrt{2}\beta}{\eta}\sqrt{1+\eta^2} \frac{\sinh{(s^{\star})}}{\cosh^3{(s^{\star})}} {\deltaMa}^{3/2} + \mathcal{O}({\deltaMa}^{5/2}),
\end{equation}
which is consistent with the amplitude of $\bar{E}$ obtained in Eq.~(\ref{Eq:Ebar_amp}) in the $\eta^2\ll\deltaMa$ limit.

\section{Displacement after $n$ reflections in a plasma slab}
\label{Sec:Infinite}

\begin{figure}
\begin{center}
\includegraphics[]{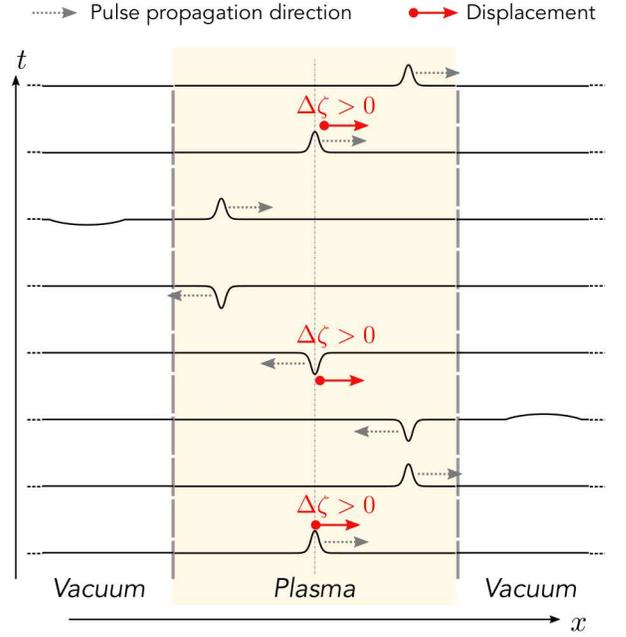}
\caption{Schematic representation of the transformation of a compressive pulse into a rarefaction pulse, and reciprocally, upon reflection at the plasma-vacuum interface in a 1d slab model. The initial conditions are those of a compressive MS soliton. The profiles depict the time evolution of the magnetic field disturbance $\delta B$. The displacement $\Delta \zeta$ due to each of the successive pulse passages adds constructively. The amplitude of the transmitted pulse is exaggerated for clarity.}
\label{Fig:Sketch_Displacement}
\end{center}
\end{figure}

With Eqs.~(\ref{Eq:displacement}, \ref{Eq:displacementDimensional}) and Eqs.~(\ref{Eq:displacement_reverse}, \ref{Eq:displacementDimensional_reverse}) in hand, we can now tackle the problem of a MS soliton propagating in a bounded plasma slab (along $\bm{\hat{x}}$). This configuration is depicted in Fig.~\ref{Fig:Sketch_Displacement}. Let us write the Mach number of the initial soliton $M_{A,0}$ and ${\deltaMa}_{,0} = M_{A,0}-1\ll1$. 

\paragraph{Pulse reflection. -- } The matching condition for the magnetic field at the plasma-vacuum interface is such that
\begin{equation}
r = \frac{B^r}{B^i} = \frac{\tilde{\kappa}^{1/2}-1}{\tilde{\kappa}^{1/2}+1},
\label{Eq:Reflection_coeff}
\end{equation} 
where, following Ref.~\cite{Heald1965}, $\mathbf{B}^i = B^i~\hat{\mathbf{z}}$ and $\mathbf{B}^r = -B^r~\hat{\mathbf{z}}$ are the magnetic field components of respectively the incident and reflected pulse. For an extraordinary wave, $\tilde{\kappa}^{1/2} = \sqrt{\varepsilon_{\perp}-{\varepsilon_{\times}}^2/\varepsilon_{\perp}}$, with $\varepsilon_{\perp}$ and $\varepsilon_{\times}$ respectively the perpendicular and cross-field component of the dielectric tensor. In the limit of low frequency waves $\omega\lesssim\omega_{ci}$, one gets $\tilde{\kappa}^{1/2}\sim\omega_{pi}/\omega_{ci}$, which can be rewritten as $\tilde{\kappa}^{1/2} = M_A/\beta$. Since $\beta/M_A\ll 1$, Eq.~(\ref{Eq:Reflection_coeff}) writes $r = 1-2\omega_{ci}/\omega_{pi}+\mathcal{O}((\beta/M_A)^2)$. With the chosen field convention, $r>0$ means that compressive pulse is thus transformed into a rarefaction pulse upon reflection at the plasma-vacuum interface, as illustrated in Fig.~\ref{Fig:Sketch_Displacement}. For an initial right propagating compressive pulse, each left propagating pulse is a rarefaction pulse, whereas each right propagating pulse is a compressive pulse. 
A consequence of this result is that the displacement $\Delta \zeta$ resulting from each successive passage of the reflected pulse adds constructively. In addition, since $2\omega_{ci}/\omega_{pi} \ll 1$, the pulse is almost entirely reflected, and only a small fraction of the incident pulse is transmitted through the interface at each interaction of the pulse with the plasma-vacuum interface.

\paragraph{Displacement from soliton bouncing. -- } In this section, it is assumed that the reflection of a soliton at the plasma vacuum-interface leads to another soliton, or, in other words, that the reflection does not change the form of the MS soliton, but only modifies its amplitude. Although rarefaction soliton solutions do not exist for transverse magnetosonic waves in cold plasma~\cite{Ohsawa2014}, it is further assumed that a rarefaction pulse such as defined in Eq.~(\ref{Eq:barBrar}) propagates with negligible change in form, \emph{i.~e.~} as a soliton. The Mach number of the $n^{th}$ reflected pulse is related to the Mach number of the $(n-1)^{th}$ reflected pulse by
\begin{equation}
M_{A,n}-1 = \left(1-2\frac{\omega_{ci}}{\omega_{pi}}\right)(M_{A,n-1}-1).
\label{Eq:Recurrence_relation}
\end{equation}
Here, use has been made of the relation $B_m = 2M_A-1$ between the soliton amplitude $B_m$ and the Mach number $M_A$. The Mach number of the $n^{th}$ reflected pulse is hence related to the Mach number of the initial pulse by
\begin{align}
{\deltaMa}_{,n} & = \left(1-2\frac{\omega_{ci}}{\omega_{pi}}\right){\deltaMa}_{,n-1}\nonumber\\ & = \left(1-2\frac{\omega_{ci}}{\omega_{pi}}\right)^n {\deltaMa}_{,0}.
\label{Eq:deltam} 
\end{align}
Using Eq.~(\ref{Eq:displacementDimensional}) and Eq.~(\ref{Eq:displacementDimensional_reverse}), the total displacement after an infinite number of reflections is
\begin{equation}
\Delta x^{\infty} = \frac{4\sqrt{2}}{1+\eta^2}\frac{c}{\omega_{pe}}\sum_{i=0}^{\infty} \left({\Xi^c}_{2i}+{\Xi^r}_{1+2i}\right)
\end{equation}
with 
\begin{subequations}
\begin{equation}
{\Xi^c}_{2i} = \sqrt{{\deltaMa}_{,2i}}\left[1 + \frac{4-3\sqrt{1+\eta^2}}{3(1+\eta^2)^{3/2}} {{\deltaMa}_{,2i}}+ \mathcal{O}\left({{\deltaMa}_{,2i}}^{2}\right)\right]
\end{equation}
\begin{multline}
{\Xi^r}_{1+2i} =\sqrt{{\deltaMa}_{,1+2i}}\left[1 - \frac{8+3\sqrt{1+\eta^2}}{3(1+\eta^2)^{3/2}} {{\deltaMa}_{,1+2i}}\right.\\\left.+ \mathcal{O}\left({{\deltaMa}_{,1+2i}}^{2}\vphantom{\frac12}\right)\right]
\end{multline}
\end{subequations}
which, using Eq.~(\ref{Eq:deltam}), and noting that
\begin{subequations}
\begin{equation}
\sum_{i=0}^{\infty}\left(1-2\frac{\omega_{ci}}{\omega_{pi}}\right)^{i} = \frac{1}{2}\frac{\omega_{pi}}{\omega_{ci}}
\end{equation}
\begin{equation}
\sum_{i=0}^{\infty}\left(1-2\frac{\omega_{ci}}{\omega_{pi}}\right)^{3i} = \frac{1}{6}\frac{\omega_{pi}}{\omega_{ci}} +\frac{1}{3}+ \mathcal{O}\left(\frac{\omega_{ci}}{\omega_{pi}}\right),
\end{equation}
\begin{equation}
\sum_{i=0}^{\infty}\left(1-2\frac{\omega_{ci}}{\omega_{pi}}\right)^{i+1/2} = \frac{1}{2}\frac{\omega_{pi}}{\omega_{ci}} -\frac{1}{2}+ \mathcal{O}\left(\frac{\omega_{ci}}{\omega_{pi}}\right),
\end{equation}
and
\begin{equation}
\sum_{i=0}^{\infty}\left(1-2\frac{\omega_{ci}}{\omega_{pi}}\right)^{\frac{3(1+2i)}{2}} = \frac{1}{6}\frac{\omega_{pi}}{\omega_{ci}} -\frac{1}{6}+ \mathcal{O}\left(\frac{\omega_{ci}}{\omega_{pi}}\right)
\end{equation}
\end{subequations}
can be written as
\begin{multline}
\Delta x^{\infty} =\frac{4\sqrt{2}}{1+\eta^2}\frac{\omega_{pi}}{\omega_{ci}}\frac{c\sqrt{{\deltaMa}_{,0}}}{\omega_{pe}} \left[1 - \frac{2+3\sqrt{1+\eta^2}}{9(1+\eta^2)^{3/2}}{{\deltaMa}_{,0}} \right.\\\left.+ \mathcal{O}\left({{\deltaMa}_{,0}}^{2}\right)\vphantom{\frac12}\right],
\label{Eq:InfiniteDisplacementDimensionalMa}
\end{multline}
or, as a function of the maximum amplitude $B_m = 1+2\deltaMa$, 
\begin{multline}
\Delta x^{\infty} =\frac{4}{1+\eta^2}\frac{\omega_{pi}}{\omega_{ci}}\frac{c}{\omega_{pe}} \sqrt{B_m-1}\\\times\left[ 1 - \frac{2+3\sqrt{1+\eta^2}}{18(1+\eta^2)^{3/2}}(B_m-1) + \mathcal{O}\left([B_m-1]^{2}\right)\right].
\label{Eq:InfiniteDisplacementDimensional}
\end{multline}
In the above expansion, the ordering $\deltaMa\gg\omega_{ci}/\omega_{pi}\gg{\deltaMa}^2$ has been assumed. For the over-dense regime, $\omega_{ci}/\omega_{pi}\ll \eta\ll1$, so that $\Delta x^{\infty}$ is larger than the electron skin depth granted that $B_m>1+\eta^2$. We then write 
\begin{align}
{\Delta x_0}^{\infty} & = \frac{4}{1+\eta^2}\frac{c}{\omega_{pe}}\frac{\omega_{pi}}{\omega_{ci}} \sqrt{B_m-1}\nonumber\\
 & = \frac{\omega_{pi}}{\omega_{ci}} {\Delta x_0}
\label{Eq:InfiniteDisplacementDimensionalZero}
\end{align}
the first order expansion of $\Delta x^{\infty}$. The displacement ${\Delta x_0}^{\infty}$ can also be written independently of the plasma density by introducing the hybrid gyro-frequency $\omega_{h} = \sqrt{\omega_{ci}\omega_{ce}}$,
\begin{equation}
{\Delta x_0}^{\infty}  = \frac{4}{1+\eta^2}\frac{c}{\omega_{h}} \sqrt{B_m-1}.
\end{equation}
In the over-dense regime considered here $\omega_h$ is also the lower-hybrid frequency $\omega_{lh} = [(\omega_{ci}\omega_{ce})^{-1}+{\omega_{pi}}^{-2}]^{-1/2}$. 

\begin{table}
\begin{center}
\begin{tabular}{c | c }
Displacement & Expression\\
\hline
Single passage & $\Delta x_0 = \frac{\displaystyle c}{\displaystyle \omega_{pe}}\frac{\displaystyle 4}{\displaystyle 1+\eta^2}\sqrt{\displaystyle B_m-1}$\\
Infinite $\#$ of passages & ${\Delta x_0}^{\infty} = \frac{\displaystyle \omega_{pi}}{\displaystyle \omega_{ci}} \Delta x_0$\\
\end{tabular}
\caption{Lowest-order expansion [$\mathcal{O}([B_m-1]^{3/2})$] of the displacement induced by a single pulse passage [Eq.~(\ref{Eq:displacementDimensionalZero})] and by an infinite number of passages after reflection in a bounded slab [Eq.~(\ref{Eq:InfiniteDisplacementDimensionalZero})].}
\label{Table:push_scaling}
\end{center}
\end{table}

\paragraph{Single ion electrostatic dynamics. -- }To validate this asymptotic development, the trajectory of a single unmagnetized ion interacting only with the the longitudinal electric field $E_x$ of the soliton pulse is simulated. At $t=0$, a longitudinal electric field 
\begin{multline}
E_x|_{t=0} = E_0\sech^2\left[\frac{\sqrt{B_m-1}}{2}\frac{\omega_{pe}}{c}(x-L/4)\right]\\\times\tanh\left[\frac{\sqrt{B_m-1}}{2}\frac{\omega_{pe}}{c}(x-L/4)\right],
\label{Eq:Electrostatic_pulse}
\end{multline}
is initialized with $E_0 = (B_m-1)^{3/2}\sqrt{m_i/m_e}V_A B_0$. This pulse propagates towards the right with a velocity $M_AV_A$. Note that compared to the model derived in Sec.~\ref{Sec:SinglePassage}, and more specifically Eqs.~(\ref{Eq:Ebar}, \ref{Eq:Ebar_amp}, \ref{Eq:Ebar_width}), the field amplitude is here smaller by a factor $M_A = (1+B_m)/2$, and the width of the pulse is larger by a factor $\sqrt{1+\eta^2}$. These choices are however consistent with standard Korteweg-de-Vries (KdV) solution (see Appendix~\ref{Sec:KdV} and Refs.~\cite{Irie2003,Ohsawa2014}). In this simple unmagnetized model, the pulse is assumed to reverse direction while maintaining form upon reaching the plasma vacuum interface. The width and amplitude of the reflected pulse are chosen as 
\begin{subequations}
\begin{equation}
w_{s}^r = w_{s}^i\left(1-2\frac{\omega_{ci}}{\omega_{pi}}\right)^{-1}
\end{equation}
\begin{equation}
E_{0}^r = E_{0}^i \left(1-2\frac{\omega_{ci}}{\omega_{pi}}\right)^{3/2}
\end{equation} 
\end{subequations}
with  $w_{s}^i$ and $E_{0}^i$ the width and amplitude of the incident pulse, respectively.

\begin{figure}
\begin{center}
\includegraphics[]{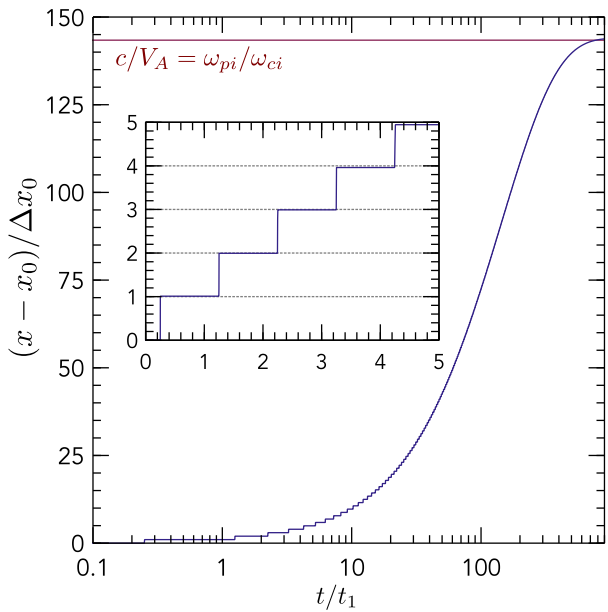}
\caption{Relative position of a test ion as a function of time as an ideal soliton is reflected successively at the plasma-vacuum interfaces of a plasma slab. Time is normalized by $t_1 = 2L\left[V_A(1+B_m)\right]^{-1}$, the transit time of a MS soliton of amplitude $B_m$ across the plasma slab of length $L$. Displacement is normalized by the first order expansion $\Delta x_0$ given in Eq.~(\ref{Eq:displacementDimensionalZero}). The first order expansion for an infinite number of reflections ${\Delta x_0}^{\infty}$, defined by Eq.~(\ref{Eq:InfiniteDisplacementDimensionalZero}), is shown in red. }
\label{Fig:InfiniteDisplacement}
\end{center}
\end{figure}

The displacement of a test ion initialized in the middle of the plasma slab of length $L_p$ ($x\in[-L_p/2,L_p/2]$) is shown in Fig.~\ref{Fig:InfiniteDisplacement} for $\omega_{ci}/\omega_{pi} = V_A/c = 7~10^{-3}$, $\eta^2 = 10^{-2}$ (\emph{i.~e.} $m_i = 100~m_e$) and $B_m-1 = 10^{-2}$ (\emph{i. e.} ${\deltaMa}_{,0} = 5~10^{-3}$). The reason for the use of a reduced ion to electron mass ratio will become clear in the next section. The computed evolution of the ion position at early times, as highlighted in the inset in Fig.~\ref{Fig:InfiniteDisplacement}, matches well the first order expansion $\Delta x_0$ given in Eq.~(\ref{Eq:displacementDimensionalZero}). A closer look confirms that the ion displacement for compressive pulses (odd displacements here) is larger than $\Delta x_0$ by about $1\%$, while it is lower than $\Delta x_0$ by about $1\%$ for rarefaction pulses (even displacements). This result is consistent with the higher order terms from Eq.~(\ref{Eq:displacementDimensional}) and Eq.~(\ref{Eq:displacementDimensional_reverse}). 


\begin{figure}
\begin{center}
\includegraphics[]{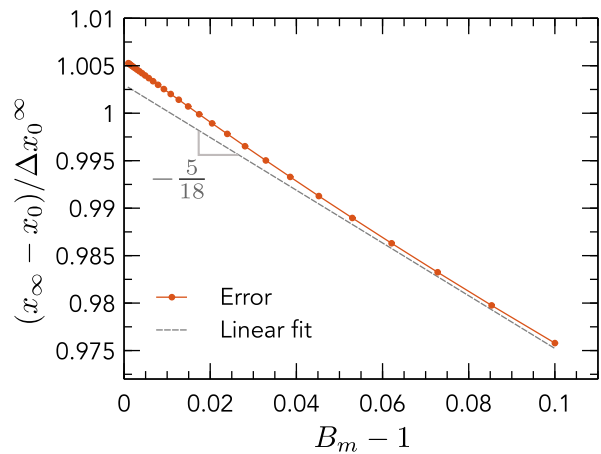}
\caption{Ratio of the ion displacement after an infinite number of  reflection, $(x_{\infty}-x_0)$, to the first order expansion ${\Delta x_0}^{\infty}$ defined by Eq.~(\ref{Eq:InfiniteDisplacementDimensionalZero}). The $B_m-1$ scaling is consistent with higher order terms in Eq.~(\ref{Eq:InfiniteDisplacementDimensional}). }
\label{Fig:Error}
\end{center}
\end{figure}

Results at long times, \emph{i.~e.} in the limit where the pulse intensity in the plasma slab goes to zero, matches well the asymptotic limit ${\Delta x_0}^{\infty}$ for the ion displacement after an infinite number of reflections derived in Eq.~(\ref{Eq:InfiniteDisplacementDimensionalZero}). This is confirmed in Fig.~\ref{Fig:Error}. The observation that the simulated displacement exceeds ${\Delta x_0}^{\infty}$ for small $\deltaMa$ can be traced back to the small differences in pulse amplitude and width discussed earlier. For stronger pulses, the linear decrease of $(x_{\infty}-x_0)/{\Delta x_0}^{\infty}$ with $B_m-1$ in Fig.~\ref{Fig:Error} is consistent with ${\deltaMa}^{3/2}$ terms in Eq.~(\ref{Eq:InfiniteDisplacementDimensional}) and the slope matches well the second order term $-(2+3\sqrt{1+\eta^2})/(18[1+\eta^2]^{3/2})\sim-5/18$.


Although this simple unmagnetized simulation confirms the asymptotic results, it is important to point out here some of its limitations. First, since the width of a soliton grows as ${\deltaMa}^{-1/2}$, the width of the ideally reflected soliton will become larger and larger as its amplitude decreases upon reflection. As a result, this model is only up to the point when the width of the soliton becomes comparable to the plasma slab width. In other words, by the $n^{th}$ reflection, the width of the soliton has grown by a factor $\varsigma$, with $\varsigma = (1-2\omega_{ci}/\omega_{pi})^{-n/2}$, while the ion displacement produced by this soliton has decreased by the same factor. In the simulation results presented above, we used $\varsigma^{-1} = 5~10^{-4}$. In addition, two strong and unphysical hypotheses of this model are the assumptions that a soliton maintains its form on reflection at the plasma-vacuum interface, and that rarefaction pulses maintain form.

\section{Numerical validation}
\label{Sec:PIC}

To relax these constraints and test the validity of the results drawn in the previous section, particle-in-cell (PIC) simulations are carried out in the same configuration.

\paragraph{Numerical model. -- } The PIC code used here is 1D version of the fully electromagnetic relativistic code \textsc{Epoch}~\cite{Arber2015}. Taking a plasma slab of a few soliton width $w_s = {\deltaMa}^{-1/2}c/\omega_{pe}$, and recalling that a small amplitude soliton ($\deltaMa\ll1$) propagates at the velocity $M_A V_A$, the simulation duration is about $\eta {M_A}^{-3/2}{\omega_{ci}}^{-1}$. Consequently, the simulation time scales roughly with the square root of the ion mass. In order to make such simulations tractable, we choose a reduced ion to electron mass ratio $\eta=10^{-2}$, and focus on the first few soliton reflections. 

To offer a valid point of comparison with the results established in the previous sections, the simulated plasma has to be cold. In particular, the peak ion kinetic energy associated with the ion longitudinal velocity in the soliton, $\varepsilon_i\sim\eta^{-2}m_e(B_m-1)^2{V_A}^2/2$, should be much larger than than the ion thermal energy. However, low plasma temperature leads to severe constraints on the number of grid points required to resolve the Debye length. As a compromise, we choose to initialize the plasma with $T_{e_0} = 0.1$~eV, $T_{i_0} = 0.03$~eV and $B_m-1 = 0.1$. The grid size is chosen equal to one Debye length. A soliton amplitude $B_m-1 = 0.1$ ensures that $\varepsilon_i$ is more than two order of magnitude larger than the ion thermal energy while remaining small enough to allow comparison with the asymptotic models for $B_m-1\ll1$ derived in the previous sections. 

The configuration simulated here consists of a plasma slab of width $L_p \sim 89 c/\omega_{pe}$, surrounded by vacuum. The total length of the simulation domain is $L=L_p/0.8$. The background magnetic field is $B_0 = 1$~T. The width of the original soliton is then of the order of $2(B_m-1)^{-1/2} c/\omega_{pe}\sim6c/\omega_{pe}$. The plasma density is $n_0 = 2~10^{21}$~m$^{-3}$, so that the over-dense regime assumption is well satisfied with  $\omega_{ce}/\omega_{pe}\sim1/14$. Accordingly, $V_A\sim c/143$ for the reduced mass $m_i=100~m_e$ used here. The pulse magnetic and electric fields, as well as ion and electron velocity fields within the pulse, are initialized in the form of a compression MS soliton (see Appendix~\ref{Sec:KdV} and Eqs.~(\ref{Eq:BzSoliton}-\ref{Eq:VyeSoliton})) located at $x=-L/6$. The main plasma parameters for the initial upstream plasma are listed in Table~\ref{Table:initial_conditions} while the PIC simulation dimensionless parameters are given in Table~\ref{Table:dimensionless_parameters}.

\begin{table}
\begin{center}
\begin{tabular}{c | c}
Parameter & Value\\
\hline
Plasma slab width $L_p$ [mm] & $10.6$\\
Ion to electron mass ratio $\eta^{-2}$ & 100\\
Electron and ion density $n_{e_0}$ and $n_{i_0}$ [cm$^{-3}$] & $2~10^{15}$\\ 
Electron temperature $T_{e_0}$ [eV] & $0.1$\\ 
Ion temperature $T_{i_0}$ [eV] & $0.03$\\ 
Background magnetic field ${B_0}$ [T] & $1$\\
Plasma frequency ${\omega_{pe_{0}}}$ [s$^{-1}$] & $2.5~10^{12}$\\
Electron gyro-frequency ${\omega_{ce_{0}}}$ [s$^{-1}$] & $1.8~10^{11}$\\
Ion gyro-frequency ${\omega_{ci_{0}}}$ [s$^{-1}$] & $1.8~10^{9}$\\
Debye length ${\lambda_{D_{0}}}$ [$\eta$m]&  $50$ \\
Electron skin depth ${c/\omega_{pe_{0}}}$ [$\mu$m]& $120$\\
Alfv{\'e}n velocity ${V_{A_0}}$ [m.s$^{-1}$] & $2.1~10^{6}$\\
Sound speed ${c_{s_0}}$ [m.s$^{-1}$] & $\sim 10^{4}$\\
\end{tabular}
\caption{Plasma upstream parameters at $t=0$. The notation $p_0$ is used to denote $p|_{t=0}$. }
\label{Table:initial_conditions}
\end{center}
\end{table}

\begin{table*}
\begin{center}
\begin{tabular}{c | c c c c c c}
Parameter & $\quad B_m-1\quad$ & $\quad\nicefrac{\displaystyle L \omega_{pe}}{\displaystyle c}\quad$ & $\quad\nicefrac{\displaystyle L_p \omega_{pe}}{\displaystyle c}\quad$ & $\quad\nicefrac{\displaystyle \omega_{pe}}{\displaystyle \omega_{ce}}\quad$ & $\quad\nicefrac{\displaystyle \omega_{pi}}{\displaystyle \omega_{ci}}\quad$ & $\quad\nicefrac{\displaystyle L_p \omega_{ci}}{\displaystyle V_A}\quad$\\
\hline
Value & $0.1$ & $112$ & $89$ & $14$ & $143$ & $9$\\
\end{tabular}
\caption{Dimensionless parameters in PIC simulations. }
\label{Table:dimensionless_parameters}
\end{center}
\end{table*}

\begin{figure*}
\begin{center}
\subfigure[~Normalized magnetic perturbation]{\includegraphics[height = 8.cm]{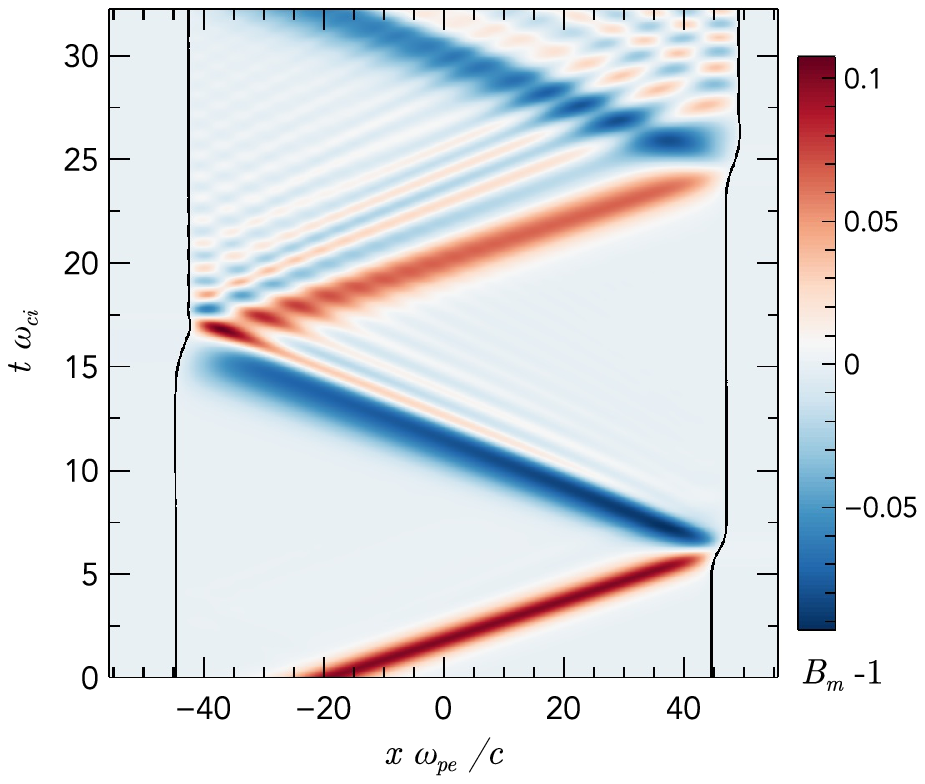}\label{Fig:BMap}}\subfigure[~Profile of the normalized magnetic field perturbation $B_m-1$ at different times]{\includegraphics[]{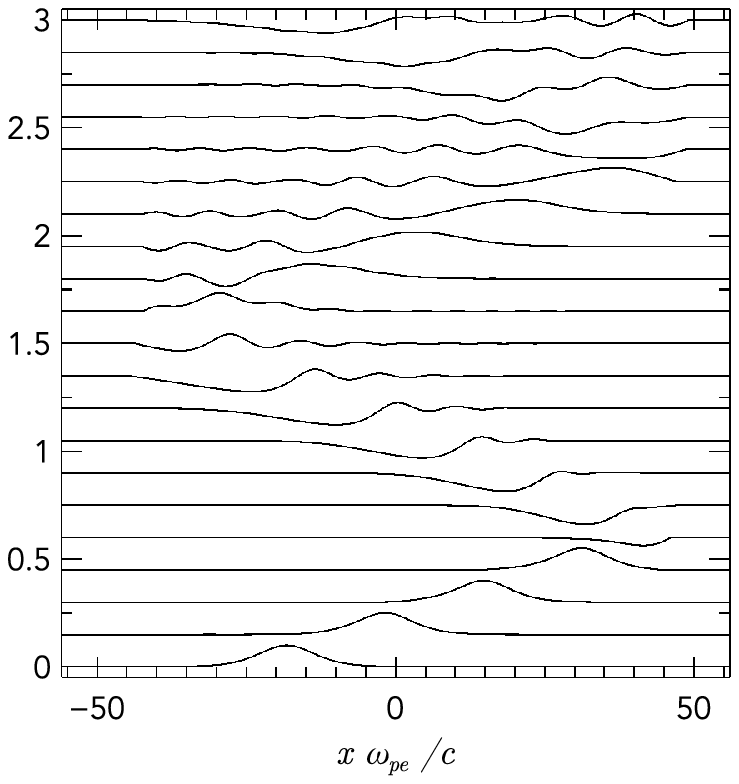}\label{Fig:StackB}}
\caption{\subref{Fig:BMap} Contour plot of the magnetic field perturbation over the entire domain, with solid-black curves denoting the plasma-vacuum boundaries, and \subref{Fig:StackB} profiles at every $\Delta t = \pi/(2\omega_{ci})$. Each profile is shifted upward by $0.15$. The self-preserving nature of the initial soliton is clearly seen until reflection ($t\omega_{ci}\sim6$) at the right plasma/vacuum boundary. Reflection leads to a rarefaction pulse and the formation of a trailing wave.}
\label{Fig:B}
\end{center}
\end{figure*}

\paragraph{Soliton propagation and reflection. -- } Fig.~\ref{Fig:B} shows the time evolution of the magnetic field over the entire computational domain. The form self-preserving property of the soliton is recovered as the compression pulse propagates towards the right up until the first reflection at the plasma-vacuum interface (plasma-vacuum boundaries are initially located near $\pm45 c/\omega_{pe}$ and depicted in dotted-black in Fig.~\ref{Fig:BMap}). Furthermore, the pulse propagation velocity inferred from the magnetic field maximum for $0\leq t\omega_{ci}\leq5$ is $1.048~V_A$, which is consistent with the phase velocity $v_{\phi} = V_A(1+B_m)/2$ of MS solitons~\cite{Adlam1958,Davis1958,Gardner1965} for the normalized pulse amplitude $B_m = 0.1$ used here.

PIC results confirm the formation of a rarefaction wave upon reflection of the compressive pulse at the interface ($t\omega_{ci}\sim6$). However, in contrast with the assumption made when computing the unmagnetized ion trajectory in the previous section, PIC results indicate that the pulse is no longer a soliton after reflection and that the left propagating pulse features a trailing wave. The result that the reflected pulse is not a soliton is analogous to what has been observed and modelled for ion-acoustic solitons~\cite{Imen1987}. As the reflected pulse propagates towards the left boundary, the amplitude of the trailing wave appears to grow. A closer look shows that the leading rarefaction pulse's width broadens while its amplitude decreases, and suggest that a fraction of the leading pulse energy is transferred to the trailing wave. A similar energy transfer from the leading pulse to the trailing wave has been reported for ion-acoustic solitons propagating in non-homogeneous unmagnetized plasmas~\cite{Ko1978,Ko1980}. This behavior is also found to result from dissipation in dispersive shock waves~\cite{Biskamp1973}.

Reflection of the left propagating rarefaction pulse and its trailing wave at the left plasma-vacuum boundary ($t\omega_{ci}\sim15$) leads to a compression pulse and a new trailing wave. Together with the first reflection, this result confirms the successive transformation of a compressive pulse into a rarefaction pulse, and vice-versa, upon reflection at the plasma-vacuum boundary. However, in contrast with the cartoon picture given in Fig.~\ref{Fig:Sketch_Displacement}, PIC simulations highlight the modifications induced by reflection on the pulse's form, and in particular the formation of a trailing wave. PIC results also indicate that the width of the leading pulse grows 

\begin{figure}
\begin{center}
\includegraphics[]{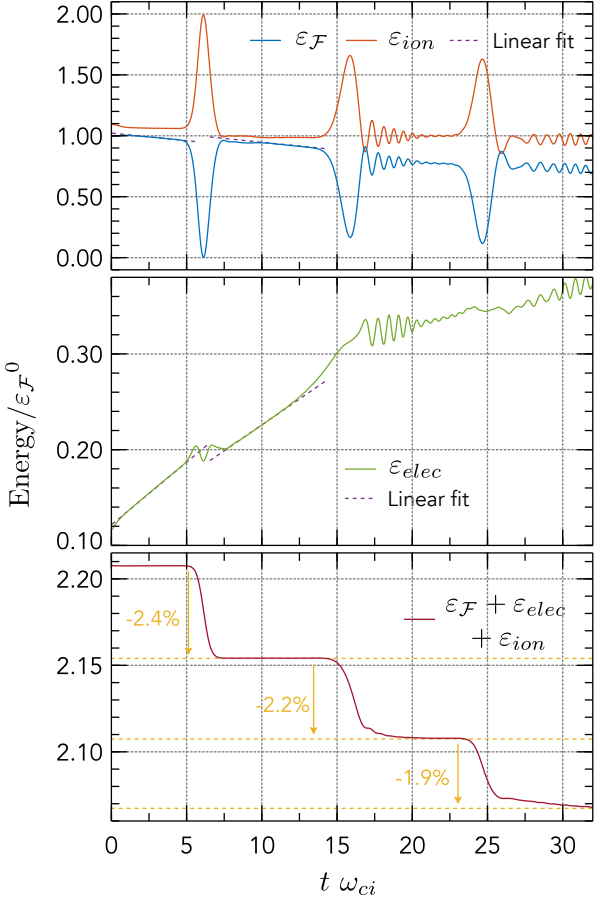}
\caption{Time evolution of the volumic energy integrated over the plasma domain for the fields ($\varepsilon_{\mathcal{F}}$, see Eq.~(\ref{Eq:SolitonFieldEnergy})), ion ($\varepsilon_{ion}$) and electron ($\varepsilon_{elec}$). Energy contents are normalized by the initial field energy content ${\varepsilon_{\mathcal{F}}}^0$ defined in Eq.~(\ref{Eq:Energy_norm}). The energy loss to the vacuum region upon pulse reflection is clearly visible on the last panel. }
\label{Fig:Energy}
\end{center}
\end{figure}

\paragraph{Pulse energy breakdown. -- } Fig.~\ref{Fig:Energy} shows the time evolution of the breakdown between the volumic energy of fields, ions and electrons integrated over the the plasma slab. The integrated field energy is defined in Eq.~(\ref{Eq:SolitonFieldEnergy}) while ions and electrons energy is obtained by summing the kinetic energy of all particle of a given species. All energies are normalized to the initial field energy content of the soliton,
\begin{equation}
{\varepsilon_\mathcal{F}}^0 = \frac{4}{3}\frac{{B_0}^2(B_m-1)^{3/2}}{\mu_0}\frac{c}{\omega_{pe}},
\label{Eq:Energy_norm}
\end{equation}
derived in Appendix~\ref{Sec:Energy}. As expected from the lowest order expansion of the KdV solution, the particles energy is initially larger than the field energy by a factor $B_m$. Other than the energy loss resulting from the transmission of part of the wave to the vacuum region upon reflection ($t\omega_{ci}\sim6$, $16$ and $24$), the last panel in Fig.~\ref{Fig:Energy} shows that the relative variation in total energy (field plus particles) in the plasma is less $10^{-4}$. The energy lost to the vacuum region as a result of the first reflection of the soliton at the plasma-vacuum interface is about $2.4\%$, that is to say that the energy reflection coefficient $R$ is about $97.6\%$. Interestingly, this figure falls in between the energy reflection coefficient 
\begin{equation}
R_l = |r|^2 = 1-4\frac{\omega_{ci}}{\omega_{pi}}+8\left(\frac{\omega_{ci}}{\omega_{pi}}\right)^2+\mathcal{O}\left(\left[\frac{\omega_{ci}}{\omega_{pi}}\right]^3\right)
\end{equation}
obtained from the continuity equation for linear waves given in Eq.~(\ref{Eq:Reflection_coeff}) and the $R_s = |r|^{3/2}$ scaling obtained by Lonngren \emph{et al.}~\cite{Lonngren1991} for KdV solitons. For the simulation parameters used here, $R_l\sim97.2\%$ and $R_s\sim97.9\%$. 

For the second and third pulse reflections ($t\omega_{ci}\sim16$) and $t\omega_{ci}\sim24$), the energy loss to the vacuum region appears to decrease slightly. The relative decrease in total energy is about $2.2\%$ and $1.9\%$, respectively. These deviations may be related to the increasing importance of radiation modes which are found in addition to the ``soliton-like'' mode. Another explanation might lie in the increase of the pulse's width. Indeed, simulations with ion-acoustic solitons showed that the reflection coefficient grows with the soliton width~\cite{Imen1987}. Notwithstanding these small deviations, the good agreement found here supports the assumption made about the amplitude of the reflected soliton when computing the unmagnetized ion trajectory in the previous section.

A remarkable feature in Fig.~\ref{Fig:Energy} is the nearly linear decrease of the pulse energy (first panel) with time in between reflections. One explanation for this behavior is energy deposition by the soliton to the electrons. More precisely, the linear decrease of $\varepsilon_{\mathcal{F}}$ is consistent with the propagation of a soliton at a velocity $M_AV_A$ and depositing an energy 
\begin{equation}
\Delta E_e = -\frac{1}{n_0 M_AV_A}\frac{d\varepsilon_{\mathcal{F}}}{dt}
\label{Eq:ElecEnergyKick}
\end{equation}
per electron. From the linear fit before the first reflection ($t\omega_{ci}\leq5$) shown in dotted-red in the first panel in Fig.~\ref{Fig:Energy}, one gets $\Delta E_e\sim0.11~$eV. This result roughly agrees with the result obtained from the slope of $\varepsilon_{elec}$ in the second panel. However, the energy deposition obtained from $\varepsilon_{elec}$ does not strictly match Eq.~(\ref{Eq:ElecEnergyKick}) since $\varepsilon_{elec}$ is a global quantity which includes phenomena occurring outside of the pulse, such as collisional effects. It is interesting to note that $\Delta E_e$ is very close to the peak longitudinal kinetic energy (see Eq.~(\ref{Eq:VxeSoliton}))
\begin{equation}
\varepsilon_{e_x} = \frac{m_e}{2}{V_A}^2(B_m-1)^2\sim 0.12~\textrm{eV},
\end{equation} 
acquired by an electron in the soliton. However, one should be cautious when trying to interpret this result. Indeed, because of the reduced mass $\eta^{-2}=100$ used here, the maximum transverse ($\varepsilon_{e_y}$) and longitudinal ($\varepsilon_{e_x}$) kinetic energy acquired by an electron in the soliton (see Eqs.~(\ref{Eq:VxeSoliton}-\ref{Eq:VyeSoliton})) only differ by a factor $4(B_m-1)\eta^{-2}/27\sim1.5$. As a result, one also has $\Delta E_e/\varepsilon_{e_y} = \mathcal{O}(1)$. Yet, simulations for a real electron to ion mass ratio should allow differentiating these two contributions since $\varepsilon_{e_y} \gg\varepsilon_{e_x}$ for $\eta^{-2}=1836$. 

A complete picture of the pulse energy breakdown between fields, electrons and ions requires considering the effects of electron-ion ($e-i$) collisions. To assess the role of collisions, one is interested in the ordering between Spitzer's equipartition time~\cite{Spitzer1962}
\begin{equation}
{\tau_{ie}}^{\varepsilon} = \frac{\displaystyle(4\pi\epsilon_0)^2}{\displaystyle 4\sqrt{2\pi}}\frac{\displaystyle \eta^{-2} \sqrt{m_e} {T_e}^{3/2}}{\displaystyle n_0 e^4 \ln{\Lambda_{ie}}},
\label{Eq:Spitzer}
\end{equation}
with $\ln{\Lambda_{ie}}$ the Coulomb logarithm, and both the soliton interaction time $\tau_r$ defined in Eq.~(\ref{Eq:PulseInteractionTime}) and the soliton propagation time $L_p/(M_AV_A)$. For the low-temperature and over-dense regime studied here, ${\tau_{ie}}^{\varepsilon}$ is a fraction of ${\omega_{ci}}^{-1}$. To the extent that $\tau_r\leq{\tau_{ie}}^{\varepsilon}$, the effect of $e-i$ collisions on the particle dynamics within the soliton can be neglected in first approximation. On the other hand, since ${\tau_{ie}}^{\varepsilon}\leq L_p/(M_AV_A)$, $e-i$ collisions will modify the plasma in between passages of the pulse. However, as discussed in Appendix~\ref{Sec:Grid_Effects}, these modifications do not appear to play a significant role on the soliton's dynamics. Furthermore, we note that since ${\tau_{ie}}^{\varepsilon}\propto \eta^{-2}$ while the simulation duration is proportional to the soliton width to soliton velocity ratio and thus scales like $\eta^{-1}$, $e-i$ collisions effects will be weaker for a real electron to ion mass ratio.

\begin{figure}
\begin{center}
\includegraphics[]{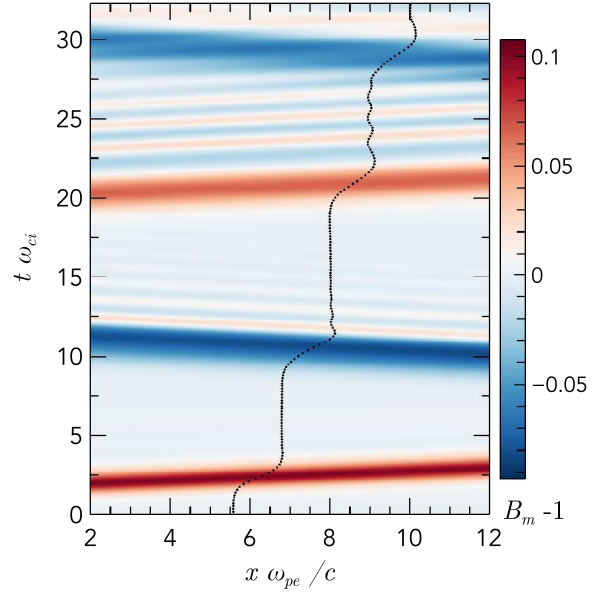}
\caption{Example of ion trajectory (in black filled-circles) predicted by PIC simulations. The trajectory is overlayed on a subset of the perturbation magnetic field map given in Fig.~\ref{Fig:BMap}. }
\label{Fig:TrajOnB}
\end{center}
\end{figure}

\paragraph{Particle displacement. -- } An example of ion trajectory obtained by averaging the PIC simulated trajectories of over $150$ individual ions initialized at $L/20\leq x_0\leq (1+10^{-4})L/20$ is overlayed on the magnetic field perturbation map in Fig.~\ref{Fig:TrajOnB}. One verifies that the passage of the pulse leads to a displacement of the particle. This displacement is in the direction of the pulse propagation for a compressive pulse, and in the direction opposite to the pulse propagation for a rarefaction pulse. Since a rarefaction pulse is turned into a compressive pulse upon reflection, and reciprocally, the displacement induced by each pulse passage adds to the previous one, as predicted in Sec.~\ref{Sec:Infinite}.

\begin{figure*}
\begin{center}
\includegraphics[]{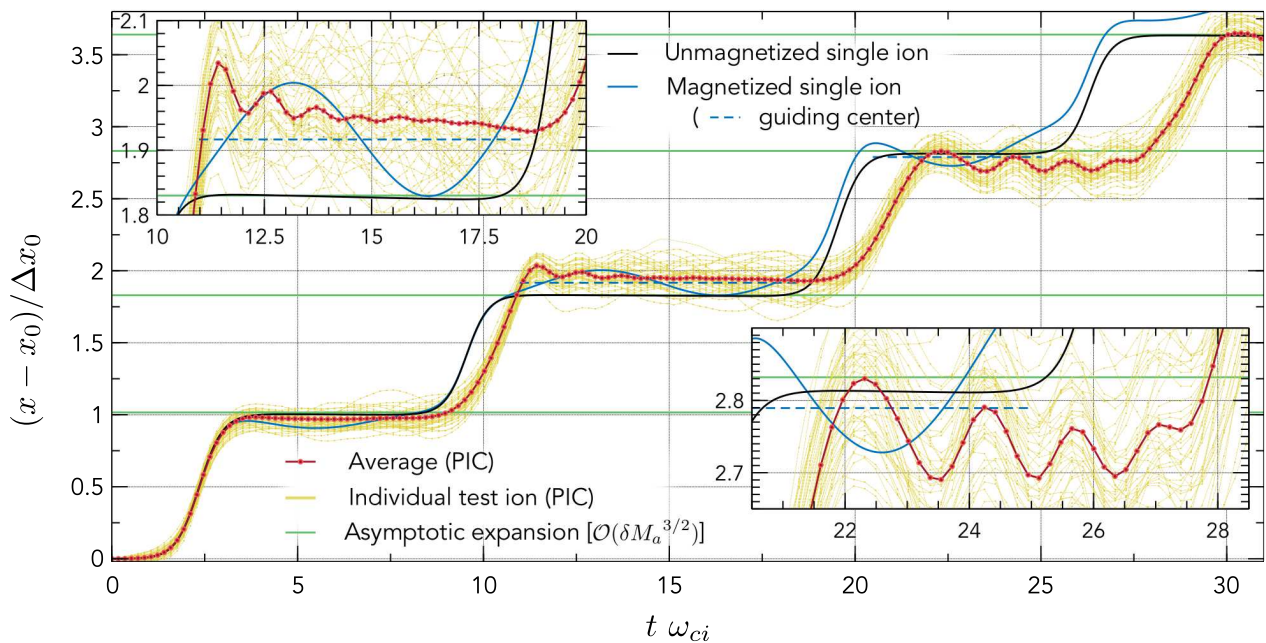}
\caption{Displacement computed for a single unmagnetized ion (see Sec.~\ref{Sec:Infinite}) and for a single magnetized ion interacting with a soliton ($E_x$, $E_y$ and $B_z$) along with the results of the electromagnetic particle-in-cell (PIC) simulation. The displacement is normalized by the first order expansion $\Delta x_0$ derived in Eq.~(\ref{Eq:displacementDimensionalZero}). Ions are located at $x_0 = L/20$ at $t=0$, and the relative variation in $x_0$ between selected PIC test ions is $\leq10^{-4}$. }
\label{Fig:PushComparison}
\end{center}
\end{figure*}

In order to quantitatively check the results derived in Sec.~\ref{Sec:SinglePassage} and Sec.~\ref{Sec:Infinite}, PIC results are compared in Fig.~\ref{Fig:PushComparison} with the asymptotic expansions Eq.~(\ref{Eq:displacementDimensional}) and Eq.~(\ref{Eq:displacementDimensional_reverse}). On this figure is also plotted the displacement obtained for an unmagnetized ion interacting with a purely electrostatic pulse ($E_x$ only) as defined in Eq.~(\ref{Eq:Electrostatic_pulse}), and for a magnetized ion interacting with an electromagnetic soliton with $E_x$, $E_y$ and $B_z$ (see Eqs.~(\ref{Eq:BzSoliton}-\ref{Eq:EySoliton})). 

The ion displacement after the first passage ($t\omega_{ci}\sim 2.5$) of the pulse agrees very well both with the unmagnetized ion model and with the asymptotic expansion $\Delta x_0$ given in Eq.~(\ref{Eq:displacementDimensionalZero}). Quantitatively, the ion displacement after the first pulse as computed from the PIC simulation is within $2\%$ of the predictions of these two idealized models. 

The second passage of the pulse as predicted by PIC results is delayed compared to single ion calculations (both magnetized and unmagnetized). This delay stems from the reflection process at the plasma-vacuum interface. As shown in the first panel in Fig.~\ref{Fig:Energy}, the soliton field energy is entirely transferred to the ions upon reaching the edge of the plasma region, before being transferred back to the field energy of the counter-propagating pulse. From Figs.~\ref{Fig:B} and~\ref{Fig:Energy}, the timescale for this energy exchange and hence for the formation of the counter-propagating pulse is $\tau_{\rightleftharpoons}\sim{\omega_{ci}}^{-1}$. On the other hand, single ion calculations assume that the pulse is immediately reflected. As a result, the second, third and fourth push predicted by PIC simulations are observed with a delay $\tau_{\rightleftharpoons}$, $2~\tau_{\rightleftharpoons}$ and $3~\tau_{\rightleftharpoons}$, respectively. 
 
The ion displacement obtained from PIC simulations after the passage of the pulse after its first reflection ($t\omega_{ci}\sim11$) is found to be about $10\%$ larger than both the asymptotic expansion to order ${\deltaMa}^{3/2}$ and the predictions of the unmagnetized ion model. This discrepancy stems from the contribution of magnetic effects for a large enough pulse amplitude. Indeed, ions acquire a transverse velocity $\mathcal{V}_y$ in response to $E_y$ in the pulse which leads to an additional displacement along $x$ by $r_L = m_p \mathcal{V}_y/(eB_0)$. Since $E_y\propto(B_m-1)$ while Eq.~(\ref{Eq:displacementDimensionalZero}) indicate that $\Delta x_0\propto \sqrt{B_m-1}$, the relative importance of this additional displacement grows with the pulse amplitude. This mechanism is confirmed by the good agreement observed between PIC results and the guiding center position obtained from the magnetized ion model once oscillations resulting from the trailing wave have faded away. 

The agreement between the asymptotic expansion and PIC results is further confirmed after the passage of the twice-reflected soliton ($t\omega_{ci}\sim21$). Quantitatively, PIC ion displacement is here found to be a few percents smaller than both the asymptotic expansion and the single ion models. One explanation for this small deviation is the energy transfer observed from the leading pulse, which accounts for most of the ion displacement, to the trailing wave as the pulse propagates across the plasma slab. Furthermore, due to the limited width of the plasma slab, the reflected rarefaction pulse begins pushing here the test ions before the trailing wave of the incident compression pulse has fully gone by.  This makes it impossible to determine the displacement after the passage of the entire right propagating pulse (leading pulse plus trailing wave).

Due to the significant computational cost of these PIC simulations, only the first three reflections of the pulse can be modeled. Nevertheless, these results confirm the main finding of this study: the passage of a MS pulse induces a displacement of particles in a magnetized plasma slab, and the displacements induced by the passages of the successive reflections of this pulse at the vacuum boundaries of a bounded plasma slab act constructively. These results also demonstrate that the reflected pulses are no longer MS solitons, similarly to what had been reported for ion-acoustic solitons~\cite{Imen1987}. Interestingly, this divergence from a pure soliton does not appear to have a strong effect on the displacement induced by those pulses, as shown by the good agreement found between PIC results and the idealized soliton reflection models. However, this observation will have to be confirmed by studying many reflections, which is beyond our current capabilities. 
 
\section{Summary}
\label{Sec:Summary}

In this paper, the plasma displacement resulting from the bouncing motion of a magnetosonic (MS) soliton within a plasma slab bounded by vacuum was investigated.

By analyzing the structure of a transverse compression MS soliton and in particular its longitudinal electric field, an analytical expression for the plasma displacement resulting from the passage of a soliton is derived in the limit of small amplitude pulses and over-dense plasmas. This displacement is roughly equal to the electron skin depth times the square root of the pulse amplitude. Then, by observing that a compression pulse is turned into a rarefaction pulse upon reflection at a plasma-vacuum boundary and vice-versa, the displacements resulting from each successive passage of a pulse bouncing back and forth in a plasma slab are shown to add up. The displacement after the pulse's energy has fully radiated to the surrounding vacuum region is found to be larger than the displacement induced by the original pulse by a factor equal to the ion plasma frequency to ion gyro-frequency ratio. This displacement is independent of the plasma density and scales as the square root of the magnetic perturbation amplitude divided by the hybrid gyro-frequency.

Particle-in-cell (PIC) simulations of the first three reflections of a compression MS soliton in a plasma slab confirm that the displacement induced by each pulse passage adds to the previous. Furthermore, PIC results corroborate the amplitude of the plasma displacement obtained from analytical models. This good agreement is particularly interesting since analytical models assume a stationary pulse form whereas PIC simulations reveal that the original soliton evolves into a pulse and a trailing wave after the first reflection. Although this agreement can only be verified here for the first three reflections, it suggests that these findings may be valid for other pulse forms.

While the plasma displacement induced by a single soliton passage is likely to be negligible for most applications, the cumulative effect associated with successive reflections may become significant for particular applications featuring over-dense plasmas.  For example, in fast magnetic compression configurations considered for plasma densification in plasma-based particle accelerators~\cite{Gueroult2016}, the soliton formed ahead of the shock~\cite{Ohsawa2017} may, under some conditions, be reflected by the density discontinuity associated with the counter-propagating shock. This mechanism would reproduce the bouncing configuration considered in this paper, and may in turn impact the plasma densification scheme. Similarly, bouncing solitons could in principle be found in between colliding shocks, both in laboratory~\cite{Dudkin1994} and space plasmas~\cite{Hietala2011}.

\section*{Acknowledgments} 
This work was supported, in part, by NNSA 67350-9960 (Prime {\#} DOE DE-NA0001836)

The authors would like to thank Dr. Laurent Garrigues for constructive discussions  and gratefully acknowledge the computing resources provided by the Princeton Plasma Physics Laboratory.

\section*{References} 

\begin{thebibliography}{66}%
\makeatletter
\providecommand \@ifxundefined [1]{%
 \@ifx{#1\undefined}
}%
\providecommand \@ifnum [1]{%
 \ifnum #1\expandafter \@firstoftwo
 \else \expandafter \@secondoftwo
 \fi
}%
\providecommand \@ifx [1]{%
 \ifx #1\expandafter \@firstoftwo
 \else \expandafter \@secondoftwo
 \fi
}%
\providecommand \natexlab [1]{#1}%
\providecommand \enquote  [1]{``#1''}%
\providecommand \bibnamefont  [1]{#1}%
\providecommand \bibfnamefont [1]{#1}%
\providecommand \citenamefont [1]{#1}%
\providecommand \href@noop [0]{\@secondoftwo}%
\providecommand \href [0]{\begingroup \@sanitize@url \@href}%
\providecommand \@href[1]{\@@startlink{#1}\@@href}%
\providecommand \@@href[1]{\endgroup#1\@@endlink}%
\providecommand \@sanitize@url [0]{\catcode `\\12\catcode `\$12\catcode
  `\&12\catcode `\#12\catcode `\^12\catcode `\_12\catcode `\%12\relax}%
\providecommand \@@startlink[1]{}%
\providecommand \@@endlink[0]{}%
\providecommand \url  [0]{\begingroup\@sanitize@url \@url }%
\providecommand \@url [1]{\endgroup\@href {#1}{\urlprefix }}%
\providecommand \urlprefix  [0]{URL }%
\providecommand \Eprint [0]{\href }%
\providecommand \doibase [0]{http://dx.doi.org/}%
\providecommand \selectlanguage [0]{\@gobble}%
\providecommand \bibinfo  [0]{\@secondoftwo}%
\providecommand \bibfield  [0]{\@secondoftwo}%
\providecommand \translation [1]{[#1]}%
\providecommand \BibitemOpen [0]{}%
\providecommand \bibitemStop [0]{}%
\providecommand \bibitemNoStop [0]{.\EOS\space}%
\providecommand \EOS [0]{\spacefactor3000\relax}%
\providecommand \BibitemShut  [1]{\csname bibitem#1\endcsname}%
\let\auto@bib@innerbib\@empty
\bibitem [{\citenamefont {Korteweg}\ and\ \citenamefont
  {de~Vries}(1895)}]{Korteweg1895}%
  \BibitemOpen
  \bibfield  {author} {\bibinfo {author} {\bibfnamefont {D.~J.}\ \bibnamefont
  {Korteweg}}\ and\ \bibinfo {author} {\bibfnamefont {G.}~\bibnamefont
  {de~Vries}},\ }\href {\doibase 10.1080/14786449508620739} {\bibfield
  {journal} {\bibinfo  {journal} {Philos. Mag.}\ }\textbf {\bibinfo {volume}
  {39}},\ \bibinfo {pages} {422} (\bibinfo {year} {1895})}\BibitemShut
  {NoStop}%
\bibitem [{\citenamefont {Wiegel}(1960)}]{Wiegel1960}%
  \BibitemOpen
  \bibfield  {author} {\bibinfo {author} {\bibfnamefont {R.~L.}\ \bibnamefont
  {Wiegel}},\ }\href {\doibase 10.1017/s0022112060001481} {\bibfield  {journal}
  {\bibinfo  {journal} {J. Fluid Mech.}\ }\textbf {\bibinfo {volume} {7}},\
  \bibinfo {pages} {273} (\bibinfo {year} {1960})}\BibitemShut {NoStop}%
\bibitem [{\citenamefont {Zabusky}\ and\ \citenamefont
  {Kruskal}(1965)}]{Zabusky1965}%
  \BibitemOpen
  \bibfield  {author} {\bibinfo {author} {\bibfnamefont {N.~J.}\ \bibnamefont
  {Zabusky}}\ and\ \bibinfo {author} {\bibfnamefont {M.~D.}\ \bibnamefont
  {Kruskal}},\ }\href {\doibase 10.1103/PhysRevLett.15.240} {\bibfield
  {journal} {\bibinfo  {journal} {Phys. Rev. Lett.}\ }\textbf {\bibinfo
  {volume} {15}},\ \bibinfo {pages} {240} (\bibinfo {year} {1965})}\BibitemShut
  {NoStop}%
\bibitem [{\citenamefont {Belashov}\ and\ \citenamefont
  {Vladimirov}(2006)}]{Belashov2006}%
  \BibitemOpen
  \bibfield  {author} {\bibinfo {author} {\bibfnamefont {V.~Y.}\ \bibnamefont
  {Belashov}}\ and\ \bibinfo {author} {\bibfnamefont {S.~V.}\ \bibnamefont
  {Vladimirov}},\ }\href {\doibase 10.1007/b138237} {\emph {\bibinfo {title}
  {Solitary Waves in Dispersive Complex Media: Theory, Simulation,
  Applications}}}\ (\bibinfo  {publisher} {Springer Berlin Heidelberg},\
  \bibinfo {year} {2006})\BibitemShut {NoStop}%
\bibitem [{\citenamefont {Washimi}\ and\ \citenamefont
  {Taniuti}(1966)}]{Washimi1966}%
  \BibitemOpen
  \bibfield  {author} {\bibinfo {author} {\bibfnamefont {H.}~\bibnamefont
  {Washimi}}\ and\ \bibinfo {author} {\bibfnamefont {T.}~\bibnamefont
  {Taniuti}},\ }\href {https://dx.doi.org/10.1103/PhysRevLett.17.996}
  {\bibfield  {journal} {\bibinfo  {journal} {Phys. Rev. Lett.}\ }\textbf
  {\bibinfo {volume} {17}},\ \bibinfo {pages} {996} (\bibinfo {year}
  {1966})}\BibitemShut {NoStop}%
\bibitem [{\citenamefont {Gardner}\ and\ \citenamefont
  {Morikawa}(1965)}]{Gardner1965}%
  \BibitemOpen
  \bibfield  {author} {\bibinfo {author} {\bibfnamefont {C.~S.}\ \bibnamefont
  {Gardner}}\ and\ \bibinfo {author} {\bibfnamefont {G.~K.}\ \bibnamefont
  {Morikawa}},\ }\href {\doibase 10.1002/cpa.3160180107} {\bibfield  {journal}
  {\bibinfo  {journal} {Commun. Pure Appl. Math.}\ }\textbf {\bibinfo {volume}
  {18}},\ \bibinfo {pages} {35} (\bibinfo {year} {1965})}\BibitemShut {NoStop}%
\bibitem [{\citenamefont {Mikhailovskii}\ and\ \citenamefont
  {Smolyakov}(1985)}]{Mikhailovskii1985}%
  \BibitemOpen
  \bibfield  {author} {\bibinfo {author} {\bibfnamefont {A.~B.}\ \bibnamefont
  {Mikhailovskii}}\ and\ \bibinfo {author} {\bibfnamefont {A.~I.}\ \bibnamefont
  {Smolyakov}},\ }\href@noop {} {\bibfield  {journal} {\bibinfo  {journal} {J.
  Exp. Theor. Phys.}\ ,\ \bibinfo {pages} {109}} (\bibinfo {year}
  {1985})}\BibitemShut {NoStop}%
\bibitem [{\citenamefont {Tran}(1974)}]{Tran1974a}%
  \BibitemOpen
  \bibfield  {author} {\bibinfo {author} {\bibfnamefont {M.~Q.}\ \bibnamefont
  {Tran}},\ }\href {\doibase 10.1088/0032-1028/16/12/006} {\bibfield  {journal}
  {\bibinfo  {journal} {Plasma Physics}\ }\textbf {\bibinfo {volume} {16}},\
  \bibinfo {pages} {1167} (\bibinfo {year} {1974})}\BibitemShut {NoStop}%
\bibitem [{\citenamefont {Ohsawa}(1985)}]{Ohsawa1985c}%
  \BibitemOpen
  \bibfield  {author} {\bibinfo {author} {\bibfnamefont {Y.}~\bibnamefont
  {Ohsawa}},\ }\href {\doibase 10.1143/jpsj.54.4073} {\bibfield  {journal}
  {\bibinfo  {journal} {J. Phys. Soc. Jpn.}\ }\textbf {\bibinfo {volume}
  {54}},\ \bibinfo {pages} {4073} (\bibinfo {year} {1985})}\BibitemShut
  {NoStop}%
\bibitem [{\citenamefont {Toida}\ and\ \citenamefont
  {Ohsawa}(1994)}]{Toida1994}%
  \BibitemOpen
  \bibfield  {author} {\bibinfo {author} {\bibfnamefont {M.}~\bibnamefont
  {Toida}}\ and\ \bibinfo {author} {\bibfnamefont {Y.}~\bibnamefont {Ohsawa}},\
  }\href {\doibase 10.1143/jpsj.63.573} {\bibfield  {journal} {\bibinfo
  {journal} {J. Phys. Soc. Jpn.}\ }\textbf {\bibinfo {volume} {63}},\ \bibinfo
  {pages} {573} (\bibinfo {year} {1994})}\BibitemShut {NoStop}%
\bibitem [{\citenamefont {Ikezi}, \citenamefont {Taylor},\ and\ \citenamefont
  {Baker}(1970)}]{Ikezi1970}%
  \BibitemOpen
  \bibfield  {author} {\bibinfo {author} {\bibfnamefont {H.}~\bibnamefont
  {Ikezi}}, \bibinfo {author} {\bibfnamefont {R.~J.}\ \bibnamefont {Taylor}}, \
  and\ \bibinfo {author} {\bibfnamefont {D.~R.}\ \bibnamefont {Baker}},\ }\href
  {\doibase 10.1103/PhysRevLett.25.11} {\bibfield  {journal} {\bibinfo
  {journal} {Phys. Rev. Lett.}\ }\textbf {\bibinfo {volume} {25}},\ \bibinfo
  {pages} {11} (\bibinfo {year} {1970})}\BibitemShut {NoStop}%
\bibitem [{\citenamefont {Stasiewicz}\ \emph {et~al.}(2003)\citenamefont
  {Stasiewicz}, \citenamefont {Shukla}, \citenamefont {Gustafsson},
  \citenamefont {Buchert}, \citenamefont {Lavraud}, \citenamefont {Thidé},\
  and\ \citenamefont {Klos}}]{Stasiewicz2003}%
  \BibitemOpen
  \bibfield  {author} {\bibinfo {author} {\bibfnamefont {K.}~\bibnamefont
  {Stasiewicz}}, \bibinfo {author} {\bibfnamefont {P.~K.}\ \bibnamefont
  {Shukla}}, \bibinfo {author} {\bibfnamefont {G.}~\bibnamefont {Gustafsson}},
  \bibinfo {author} {\bibfnamefont {S.}~\bibnamefont {Buchert}}, \bibinfo
  {author} {\bibfnamefont {B.}~\bibnamefont {Lavraud}}, \bibinfo {author}
  {\bibfnamefont {B.}~\bibnamefont {Thidé}}, \ and\ \bibinfo {author}
  {\bibfnamefont {Z.}~\bibnamefont {Klos}},\ }\href {\doibase
  10.1103/PhysRevLett.90.085002} {\bibfield  {journal} {\bibinfo  {journal}
  {Phys. Rev. Lett.}\ }\textbf {\bibinfo {volume} {90}},\ \bibinfo {pages}
  {085002} (\bibinfo {year} {2003})}\BibitemShut {NoStop}%
\bibitem [{\citenamefont {Sagdeev}(1966)}]{Sagdeev1966}%
  \BibitemOpen
  \bibfield  {author} {\bibinfo {author} {\bibfnamefont {R.~Z.}\ \bibnamefont
  {Sagdeev}},\ }\href@noop {} {\bibfield  {journal} {\bibinfo  {journal} {Rev.
  Plasma Physics}\ }\textbf {\bibinfo {volume} {4}},\ \bibinfo {pages} {23}
  (\bibinfo {year} {1966})}\BibitemShut {NoStop}%
\bibitem [{\citenamefont {Tidman}\ and\ \citenamefont
  {Krall}(1971)}]{Tidman1971}%
  \BibitemOpen
  \bibfield  {author} {\bibinfo {author} {\bibfnamefont {D.~A.}\ \bibnamefont
  {Tidman}}\ and\ \bibinfo {author} {\bibfnamefont {N.~A.}\ \bibnamefont
  {Krall}},\ }\href@noop {} {\emph {\bibinfo {title} {Shock waves in
  collisionless plasmas}}}\ (\bibinfo  {publisher} {Wiley-Interscience},\
  \bibinfo {year} {1971})\BibitemShut {NoStop}%
\bibitem [{\citenamefont {Biskamp}(1973)}]{Biskamp1973}%
  \BibitemOpen
  \bibfield  {author} {\bibinfo {author} {\bibfnamefont {D.}~\bibnamefont
  {Biskamp}},\ }\href {\doibase 10.1088/0029-5515/13/5/010} {\bibfield
  {journal} {\bibinfo  {journal} {Nucl. Fusion}\ }\textbf {\bibinfo {volume}
  {13}},\ \bibinfo {pages} {719} (\bibinfo {year} {1973})}\BibitemShut
  {NoStop}%
\bibitem [{\citenamefont {Balogh}\ and\ \citenamefont
  {Treumann}(2013)}]{Balogh2013}%
  \BibitemOpen
  \bibfield  {author} {\bibinfo {author} {\bibfnamefont {A.}~\bibnamefont
  {Balogh}}\ and\ \bibinfo {author} {\bibfnamefont {R.~A.}\ \bibnamefont
  {Treumann}},\ }\href {\doibase 10.1007/978-1-4614-6099-2} {\emph {\bibinfo
  {title} {Physics of Collisionless Shocks}}},\ \bibinfo {series} {ISSI
  Scientific Report Series}, Vol.~\bibinfo {volume} {12}\ (\bibinfo
  {publisher} {Springer-Verlag New York},\ \bibinfo {year} {2013})\BibitemShut
  {NoStop}%
\bibitem [{\citenamefont {Ohsawa}(2014)}]{Ohsawa2014}%
  \BibitemOpen
  \bibfield  {author} {\bibinfo {author} {\bibfnamefont {Y.}~\bibnamefont
  {Ohsawa}},\ }\href {\doibase 10.1016/j.physrep.2013.11.004} {\bibfield
  {journal} {\bibinfo  {journal} {Phys. Rep.}\ }\textbf {\bibinfo {volume}
  {536}},\ \bibinfo {pages} {147} (\bibinfo {year} {2014})}\BibitemShut
  {NoStop}%
\bibitem [{\citenamefont {Gueroult}, \citenamefont {Ohsawa},\ and\
  \citenamefont {Fisch}(2017)}]{Gueroult2017}%
  \BibitemOpen
  \bibfield  {author} {\bibinfo {author} {\bibfnamefont {R.}~\bibnamefont
  {Gueroult}}, \bibinfo {author} {\bibfnamefont {Y.}~\bibnamefont {Ohsawa}}, \
  and\ \bibinfo {author} {\bibfnamefont {N.~J.}\ \bibnamefont {Fisch}},\ }\href
  {\doibase 10.1103/PhysRevLett.118.125101} {\bibfield  {journal} {\bibinfo
  {journal} {Phys. Rev. Lett.}\ }\textbf {\bibinfo {volume} {118}},\ \bibinfo
  {pages} {125101} (\bibinfo {year} {2017})}\BibitemShut {NoStop}%
\bibitem [{\citenamefont {Davis}, \citenamefont {Lust},\ and\ \citenamefont
  {Schluter}(1958)}]{Davis1958}%
  \BibitemOpen
  \bibfield  {author} {\bibinfo {author} {\bibfnamefont {L.}~\bibnamefont
  {Davis}}, \bibinfo {author} {\bibfnamefont {R.}~\bibnamefont {Lust}}, \ and\
  \bibinfo {author} {\bibfnamefont {A.}~\bibnamefont {Schluter}},\ }\href
  {http://zfn.mpdl.mpg.de/data/Reihe_A/13/ZNA-1958-13a- 0916.pdf} {\bibfield
  {journal} {\bibinfo  {journal} {Z. Naturforsch A}\ }\textbf {\bibinfo
  {volume} {13a}},\ \bibinfo {pages} {916} (\bibinfo {year}
  {1958})}\BibitemShut {NoStop}%
\bibitem [{\citenamefont {Adlam}\ and\ \citenamefont
  {Allen}(1958)}]{Adlam1958}%
  \BibitemOpen
  \bibfield  {author} {\bibinfo {author} {\bibfnamefont {J.~H.}\ \bibnamefont
  {Adlam}}\ and\ \bibinfo {author} {\bibfnamefont {J.~E.}\ \bibnamefont
  {Allen}},\ }\href {\doibase 10.1080/14786435808244566} {\bibfield  {journal}
  {\bibinfo  {journal} {Philos. Mag.}\ }\textbf {\bibinfo {volume} {3}},\
  \bibinfo {pages} {448} (\bibinfo {year} {1958})}\BibitemShut {NoStop}%
\bibitem [{\citenamefont {Adlam}\ and\ \citenamefont
  {Allen}(1960)}]{Adlam1960}%
  \BibitemOpen
  \bibfield  {author} {\bibinfo {author} {\bibfnamefont {J.~H.}\ \bibnamefont
  {Adlam}}\ and\ \bibinfo {author} {\bibfnamefont {J.~E.}\ \bibnamefont
  {Allen}},\ }\href {\doibase 10.1088/0370-1328/75/5/302} {\bibfield  {journal}
  {\bibinfo  {journal} {Proc. Phys. Soc.}\ }\textbf {\bibinfo {volume} {75}},\
  \bibinfo {pages} {640} (\bibinfo {year} {1960})}\BibitemShut {NoStop}%
\bibitem [{\citenamefont {Ba{\~n}os}\ and\ \citenamefont
  {Vernon}(1960)}]{Banos1960}%
  \BibitemOpen
  \bibfield  {author} {\bibinfo {author} {\bibfnamefont {A.}~\bibnamefont
  {Ba{\~n}os}}\ and\ \bibinfo {author} {\bibfnamefont {R.}~\bibnamefont
  {Vernon}},\ }\href {\doibase 10.1007/BF02860251} {\bibfield  {journal}
  {\bibinfo  {journal} {Il Nuovo Cimento}\ }\textbf {\bibinfo {volume} {15}},\
  \bibinfo {pages} {269} (\bibinfo {year} {1960})}\BibitemShut {NoStop}%
\bibitem [{\citenamefont {Hain}, \citenamefont {Lüst},\ and\ \citenamefont
  {Schlüter}(1960)}]{Hain1960}%
  \BibitemOpen
  \bibfield  {author} {\bibinfo {author} {\bibfnamefont {K.}~\bibnamefont
  {Hain}}, \bibinfo {author} {\bibfnamefont {R.}~\bibnamefont {Lüst}}, \ and\
  \bibinfo {author} {\bibfnamefont {A.}~\bibnamefont {Schlüter}},\ }\href
  {\doibase 10.1103/RevModPhys.32.967} {\bibfield  {journal} {\bibinfo
  {journal} {Rev. Mod. Phys.}\ }\textbf {\bibinfo {volume} {32}},\ \bibinfo
  {pages} {967} (\bibinfo {year} {1960})}\BibitemShut {NoStop}%
\bibitem [{\citenamefont {Ohsawa}(1986)}]{Ohsawa1986a}%
  \BibitemOpen
  \bibfield  {author} {\bibinfo {author} {\bibfnamefont {Y.}~\bibnamefont
  {Ohsawa}},\ }\href {\doibase 10.1063/1.865614} {\bibfield  {journal}
  {\bibinfo  {journal} {Phys. Fluids}\ }\textbf {\bibinfo {volume} {29}},\
  \bibinfo {pages} {1844} (\bibinfo {year} {1986})}\BibitemShut {NoStop}%
\bibitem [{\citenamefont {Ohsawa}(1990)}]{Ohsawa1990}%
  \BibitemOpen
  \bibfield  {author} {\bibinfo {author} {\bibfnamefont {Y.}~\bibnamefont
  {Ohsawa}},\ }\href {\doibase 10.1143/JPSJ.59.2782} {\bibfield  {journal}
  {\bibinfo  {journal} {J. Phys. Soc. Jpn.}\ }\textbf {\bibinfo {volume}
  {59}},\ \bibinfo {pages} {2782} (\bibinfo {year} {1990})}\BibitemShut
  {NoStop}%
\bibitem [{\citenamefont {Rau}\ and\ \citenamefont {Tajima}(1998)}]{Rau1998}%
  \BibitemOpen
  \bibfield  {author} {\bibinfo {author} {\bibfnamefont {B.}~\bibnamefont
  {Rau}}\ and\ \bibinfo {author} {\bibfnamefont {T.}~\bibnamefont {Tajima}},\
  }\href {\doibase 10.1063/1.873076} {\bibfield  {journal} {\bibinfo  {journal}
  {Phys. Plasmas}\ }\textbf {\bibinfo {volume} {5}},\ \bibinfo {pages} {3575}
  (\bibinfo {year} {1998})}\BibitemShut {NoStop}%
\bibitem [{\citenamefont {Maruyama}, \citenamefont {Bessho},\ and\
  \citenamefont {Ohsawa}(1998)}]{Maruyama1998}%
  \BibitemOpen
  \bibfield  {author} {\bibinfo {author} {\bibfnamefont {K.}~\bibnamefont
  {Maruyama}}, \bibinfo {author} {\bibfnamefont {N.}~\bibnamefont {Bessho}}, \
  and\ \bibinfo {author} {\bibfnamefont {Y.}~\bibnamefont {Ohsawa}},\ }\href
  {\doibase 10.1063/1.872993} {\bibfield  {journal} {\bibinfo  {journal} {Phys.
  Plasmas}\ }\textbf {\bibinfo {volume} {5}},\ \bibinfo {pages} {3257}
  (\bibinfo {year} {1998})}\BibitemShut {NoStop}%
\bibitem [{\citenamefont {Stasiewicz}(2007)}]{Stasiewicz2007}%
  \BibitemOpen
  \bibfield  {author} {\bibinfo {author} {\bibfnamefont {K.}~\bibnamefont
  {Stasiewicz}},\ }\href {\doibase 10.1088/0741-3335/49/12B/S58} {\bibfield
  {journal} {\bibinfo  {journal} {Plasma Phys. Controlled Fusion}\ }\textbf
  {\bibinfo {volume} {49}},\ \bibinfo {pages} {B621} (\bibinfo {year}
  {2007})}\BibitemShut {NoStop}%
\bibitem [{\citenamefont {Ohsawa}(2017{\natexlab{a}})}]{Ohsawa2017a}%
  \BibitemOpen
  \bibfield  {author} {\bibinfo {author} {\bibfnamefont {Y.}~\bibnamefont
  {Ohsawa}},\ }\href {\doibase 10.1063/1.5001997} {\bibfield  {journal}
  {\bibinfo  {journal} {Phys. Plasmas}\ }\textbf {\bibinfo {volume} {24}},\
  \bibinfo {pages} {112304} (\bibinfo {year} {2017}{\natexlab{a}})}\BibitemShut
  {NoStop}%
\bibitem [{\citenamefont {Lembege}\ \emph {et~al.}(1983)\citenamefont
  {Lembege}, \citenamefont {Ratliff}, \citenamefont {Dawson},\ and\
  \citenamefont {Ohsawa}}]{Lembege1983}%
  \BibitemOpen
  \bibfield  {author} {\bibinfo {author} {\bibfnamefont {B.}~\bibnamefont
  {Lembege}}, \bibinfo {author} {\bibfnamefont {S.~T.}\ \bibnamefont
  {Ratliff}}, \bibinfo {author} {\bibfnamefont {J.~M.}\ \bibnamefont {Dawson}},
  \ and\ \bibinfo {author} {\bibfnamefont {Y.}~\bibnamefont {Ohsawa}},\ }\href
  {\doibase 10.1103/PhysRevLett.51.264} {\bibfield  {journal} {\bibinfo
  {journal} {Phys. Rev. Lett.}\ }\textbf {\bibinfo {volume} {51}},\ \bibinfo
  {pages} {264} (\bibinfo {year} {1983})}\BibitemShut {NoStop}%
\bibitem [{\citenamefont {Lembege}\ and\ \citenamefont
  {Dawson}(1984)}]{Lembege1984}%
  \BibitemOpen
  \bibfield  {author} {\bibinfo {author} {\bibfnamefont {B.}~\bibnamefont
  {Lembege}}\ and\ \bibinfo {author} {\bibfnamefont {J.~M.}\ \bibnamefont
  {Dawson}},\ }\href {\doibase 10.1103/PhysRevLett.53.1053} {\bibfield
  {journal} {\bibinfo  {journal} {Phys. Rev. Lett.}\ }\textbf {\bibinfo
  {volume} {53}},\ \bibinfo {pages} {1053} (\bibinfo {year}
  {1984})}\BibitemShut {NoStop}%
\bibitem [{\citenamefont {Lembege}\ and\ \citenamefont
  {Dawson}(1986)}]{Lembege1986}%
  \BibitemOpen
  \bibfield  {author} {\bibinfo {author} {\bibfnamefont {B.}~\bibnamefont
  {Lembege}}\ and\ \bibinfo {author} {\bibfnamefont {J.~M.}\ \bibnamefont
  {Dawson}},\ }\href {\doibase 10.1063/1.865939} {\bibfield  {journal}
  {\bibinfo  {journal} {Phys. Fluids}\ }\textbf {\bibinfo {volume} {29}},\
  \bibinfo {pages} {821} (\bibinfo {year} {1986})}\BibitemShut {NoStop}%
\bibitem [{\citenamefont {Lembege}\ and\ \citenamefont
  {Dawson}(1989)}]{Lembege1989}%
  \BibitemOpen
  \bibfield  {author} {\bibinfo {author} {\bibfnamefont {B.}~\bibnamefont
  {Lembege}}\ and\ \bibinfo {author} {\bibfnamefont {J.~M.}\ \bibnamefont
  {Dawson}},\ }\href {\doibase 10.1063/1.859021} {\bibfield  {journal}
  {\bibinfo  {journal} {Phys. Fluids B: Plasma Phys.}\ }\textbf {\bibinfo
  {volume} {1}},\ \bibinfo {pages} {1001} (\bibinfo {year} {1989})}\BibitemShut
  {NoStop}%
\bibitem [{\citenamefont {Drury}(1983)}]{Drury1983}%
  \BibitemOpen
  \bibfield  {author} {\bibinfo {author} {\bibfnamefont {L.~O.}\ \bibnamefont
  {Drury}},\ }\href {http://stacks.iop.org/0034-4885/46/i=8/a=002} {\bibfield
  {journal} {\bibinfo  {journal} {Rep. Prog. Phys.}\ }\textbf {\bibinfo
  {volume} {46}},\ \bibinfo {pages} {973} (\bibinfo {year} {1983})}\BibitemShut
  {NoStop}%
\bibitem [{\citenamefont {Yuan}\ \emph {et~al.}(2013)\citenamefont {Yuan},
  \citenamefont {Xiao}, \citenamefont {Luo}, \citenamefont {Walker},
  \citenamefont {Welander}, \citenamefont {Hyatt}, \citenamefont {Qian},
  \citenamefont {Zhang}, \citenamefont {Humphreys}, \citenamefont {Leuer},
  \citenamefont {Johnson}, \citenamefont {Penaflor},\ and\ \citenamefont
  {Mueller}}]{Yuan2013}%
  \BibitemOpen
  \bibfield  {author} {\bibinfo {author} {\bibfnamefont {Q.~P.}\ \bibnamefont
  {Yuan}}, \bibinfo {author} {\bibfnamefont {B.~J.}\ \bibnamefont {Xiao}},
  \bibinfo {author} {\bibfnamefont {Z.~P.}\ \bibnamefont {Luo}}, \bibinfo
  {author} {\bibfnamefont {M.~L.}\ \bibnamefont {Walker}}, \bibinfo {author}
  {\bibfnamefont {A.~S.}\ \bibnamefont {Welander}}, \bibinfo {author}
  {\bibfnamefont {A.}~\bibnamefont {Hyatt}}, \bibinfo {author} {\bibfnamefont
  {J.~P.}\ \bibnamefont {Qian}}, \bibinfo {author} {\bibfnamefont {R.~R.}\
  \bibnamefont {Zhang}}, \bibinfo {author} {\bibfnamefont {D.~A.}\ \bibnamefont
  {Humphreys}}, \bibinfo {author} {\bibfnamefont {J.~A.}\ \bibnamefont
  {Leuer}}, \bibinfo {author} {\bibfnamefont {R.~D.}\ \bibnamefont {Johnson}},
  \bibinfo {author} {\bibfnamefont {B.~G.}\ \bibnamefont {Penaflor}}, \ and\
  \bibinfo {author} {\bibfnamefont {D.}~\bibnamefont {Mueller}},\ }\href
  {\doibase 10.1088/0029-5515/53/4/043009} {\bibfield  {journal} {\bibinfo
  {journal} {Nucl. Fusion}\ }\textbf {\bibinfo {volume} {53}},\ \bibinfo
  {pages} {043009} (\bibinfo {year} {2013})}\BibitemShut {NoStop}%
\bibitem [{\citenamefont {Canali}\ \emph {et~al.}(2011)\citenamefont {Canali},
  \citenamefont {Carraro}, \citenamefont {Krasnicky}, \citenamefont
  {Lagomarsino}, \citenamefont {Di~Noto}, \citenamefont {Testera},\ and\
  \citenamefont {Zavatarelli}}]{Canali2011}%
  \BibitemOpen
  \bibfield  {author} {\bibinfo {author} {\bibfnamefont {C.}~\bibnamefont
  {Canali}}, \bibinfo {author} {\bibfnamefont {C.}~\bibnamefont {Carraro}},
  \bibinfo {author} {\bibfnamefont {D.}~\bibnamefont {Krasnicky}}, \bibinfo
  {author} {\bibfnamefont {V.}~\bibnamefont {Lagomarsino}}, \bibinfo {author}
  {\bibfnamefont {L.}~\bibnamefont {Di~Noto}}, \bibinfo {author} {\bibfnamefont
  {G.}~\bibnamefont {Testera}}, \ and\ \bibinfo {author} {\bibfnamefont
  {S.}~\bibnamefont {Zavatarelli}},\ }\href {\doibase
  10.1140/epjd/e2011-20552-x} {\bibfield  {journal} {\bibinfo  {journal} {Eur.
  Phys. J. D}\ }\textbf {\bibinfo {volume} {65}},\ \bibinfo {pages} {499}
  (\bibinfo {year} {2011})}\BibitemShut {NoStop}%
\bibitem [{\citenamefont {Lifshitz}\ \emph {et~al.}(2012)\citenamefont
  {Lifshitz}, \citenamefont {Be'ery}, \citenamefont {Fisher}, \citenamefont
  {Ron},\ and\ \citenamefont {Fruchtman}}]{Lifshitz2012}%
  \BibitemOpen
  \bibfield  {author} {\bibinfo {author} {\bibfnamefont {A.}~\bibnamefont
  {Lifshitz}}, \bibinfo {author} {\bibfnamefont {I.}~\bibnamefont {Be'ery}},
  \bibinfo {author} {\bibfnamefont {A.}~\bibnamefont {Fisher}}, \bibinfo
  {author} {\bibfnamefont {A.}~\bibnamefont {Ron}}, \ and\ \bibinfo {author}
  {\bibfnamefont {A.}~\bibnamefont {Fruchtman}},\ }\href {\doibase
  10.1088/0029-5515/52/12/123008} {\bibfield  {journal} {\bibinfo  {journal}
  {Nucl. Fusion}\ }\textbf {\bibinfo {volume} {52}},\ \bibinfo {pages} {123008}
  (\bibinfo {year} {2012})}\BibitemShut {NoStop}%
\bibitem [{\citenamefont {Lieberman}\ and\ \citenamefont
  {Lichtenberg}(1994)}]{Lieberman1994}%
  \BibitemOpen
  \bibfield  {author} {\bibinfo {author} {\bibfnamefont {M.~A.}\ \bibnamefont
  {Lieberman}}\ and\ \bibinfo {author} {\bibfnamefont {A.~J.}\ \bibnamefont
  {Lichtenberg}},\ }\href@noop {} {\emph {\bibinfo {title} {Principles of
  Plasma Discharge for Materials Processing}}}\ (\bibinfo  {publisher} {John
  Wiley \& Sons},\ \bibinfo {year} {1994})\ p.\ \bibinfo {pages}
  {600}\BibitemShut {NoStop}%
\bibitem [{\citenamefont {Taniuti}\ and\ \citenamefont
  {Wei}(1968)}]{Taniuti1968}%
  \BibitemOpen
  \bibfield  {author} {\bibinfo {author} {\bibfnamefont {T.}~\bibnamefont
  {Taniuti}}\ and\ \bibinfo {author} {\bibfnamefont {C.-C.}\ \bibnamefont
  {Wei}},\ }\href {\doibase 10.1143/jpsj.24.941} {\bibfield  {journal}
  {\bibinfo  {journal} {J. Phys. Soc. Jpn.}\ }\textbf {\bibinfo {volume}
  {24}},\ \bibinfo {pages} {941} (\bibinfo {year} {1968})}\BibitemShut
  {NoStop}%
\bibitem [{\citenamefont {Ko}\ and\ \citenamefont {Kuehl}(1978)}]{Ko1978}%
  \BibitemOpen
  \bibfield  {author} {\bibinfo {author} {\bibfnamefont {K.}~\bibnamefont
  {Ko}}\ and\ \bibinfo {author} {\bibfnamefont {H.~H.}\ \bibnamefont {Kuehl}},\
  }\href {\doibase 10.1103/PhysRevLett.40.233} {\bibfield  {journal} {\bibinfo
  {journal} {Phys. Rev. Lett.}\ }\textbf {\bibinfo {volume} {40}},\ \bibinfo
  {pages} {233} (\bibinfo {year} {1978})}\BibitemShut {NoStop}%
\bibitem [{\citenamefont {Kakutani}(1971)}]{Kakutani1971}%
  \BibitemOpen
  \bibfield  {author} {\bibinfo {author} {\bibfnamefont {T.}~\bibnamefont
  {Kakutani}},\ }\href {\doibase 10.1143/jpsj.31.1246} {\bibfield  {journal}
  {\bibinfo  {journal} {J. Phys. Soc. Jpn.}\ }\textbf {\bibinfo {volume}
  {31}},\ \bibinfo {pages} {1246} (\bibinfo {year} {1971})}\BibitemShut
  {NoStop}%
\bibitem [{\citenamefont {Berezin}\ and\ \citenamefont
  {Karpman}(1967)}]{Berezin1967}%
  \BibitemOpen
  \bibfield  {author} {\bibinfo {author} {\bibfnamefont {Y.~A.}\ \bibnamefont
  {Berezin}}\ and\ \bibinfo {author} {\bibfnamefont {V.~I.}\ \bibnamefont
  {Karpman}},\ }\href@noop {} {\bibfield  {journal} {\bibinfo  {journal} {J.
  Exp. Theor. Phys.}\ }\textbf {\bibinfo {volume} {24}},\ \bibinfo {pages}
  {1049} (\bibinfo {year} {1967})}\BibitemShut {NoStop}%
\bibitem [{\citenamefont {Tappert}\ and\ \citenamefont
  {Zabusky}(1971)}]{Tappert1971}%
  \BibitemOpen
  \bibfield  {author} {\bibinfo {author} {\bibfnamefont {F.~D.}\ \bibnamefont
  {Tappert}}\ and\ \bibinfo {author} {\bibfnamefont {N.~J.}\ \bibnamefont
  {Zabusky}},\ }\href {http://link.aps.org/doi/10.1103/PhysRevLett.27.1774}
  {\bibfield  {journal} {\bibinfo  {journal} {Phys. Rev. Lett.}\ }\textbf
  {\bibinfo {volume} {27}},\ \bibinfo {pages} {1774} (\bibinfo {year}
  {1971})}\BibitemShut {NoStop}%
\bibitem [{\citenamefont {Karpman}\ and\ \citenamefont
  {Maslov}(1977)}]{Karpman1977}%
  \BibitemOpen
  \bibfield  {author} {\bibinfo {author} {\bibfnamefont {V.~I.}\ \bibnamefont
  {Karpman}}\ and\ \bibinfo {author} {\bibfnamefont {E.~M.}\ \bibnamefont
  {Maslov}},\ }\href@noop {} {\bibfield  {journal} {\bibinfo  {journal} {J.
  Exp. Theor. Phys.}\ }\textbf {\bibinfo {volume} {46}},\ \bibinfo {pages}
  {281} (\bibinfo {year} {1977})}\BibitemShut {NoStop}%
\bibitem [{\citenamefont {Karpman}\ and\ \citenamefont
  {Maslov}(1978)}]{Karpman1978}%
  \BibitemOpen
  \bibfield  {author} {\bibinfo {author} {\bibfnamefont {V.~I.}\ \bibnamefont
  {Karpman}}\ and\ \bibinfo {author} {\bibfnamefont {E.~M.}\ \bibnamefont
  {Maslov}},\ }\href@noop {} {\bibfield  {journal} {\bibinfo  {journal} {J.
  Exp. Theor. Phys.}\ }\textbf {\bibinfo {volume} {48}},\ \bibinfo {pages}
  {252} (\bibinfo {year} {1978})}\BibitemShut {NoStop}%
\bibitem [{\citenamefont {Ko}\ and\ \citenamefont {Kuehl}(1980)}]{Ko1980}%
  \BibitemOpen
  \bibfield  {author} {\bibinfo {author} {\bibfnamefont {K.}~\bibnamefont
  {Ko}}\ and\ \bibinfo {author} {\bibfnamefont {H.~H.}\ \bibnamefont {Kuehl}},\
  }\href {\doibase 10.1063/1.863065} {\bibfield  {journal} {\bibinfo  {journal}
  {Phys. Fluids}\ }\textbf {\bibinfo {volume} {23}},\ \bibinfo {pages} {834}
  (\bibinfo {year} {1980})}\BibitemShut {NoStop}%
\bibitem [{\citenamefont {Nakata}(1988)}]{Nakata1988}%
  \BibitemOpen
  \bibfield  {author} {\bibinfo {author} {\bibfnamefont {I.}~\bibnamefont
  {Nakata}},\ }\href {\doibase 10.1143/jpsj.57.2209} {\bibfield  {journal}
  {\bibinfo  {journal} {J. Phys. Soc. Jpn.}\ }\textbf {\bibinfo {volume}
  {57}},\ \bibinfo {pages} {2209} (\bibinfo {year} {1988})}\BibitemShut
  {NoStop}%
\bibitem [{\citenamefont {Lonngren}, \citenamefont {Andersen},\ and\
  \citenamefont {Cooney}(1991)}]{Lonngren1991}%
  \BibitemOpen
  \bibfield  {author} {\bibinfo {author} {\bibfnamefont {K.~E.}\ \bibnamefont
  {Lonngren}}, \bibinfo {author} {\bibfnamefont {D.~R.}\ \bibnamefont
  {Andersen}}, \ and\ \bibinfo {author} {\bibfnamefont {J.~L.}\ \bibnamefont
  {Cooney}},\ }\href {\doibase 10.1016/0375-9601(91)90724-M} {\bibfield
  {journal} {\bibinfo  {journal} {Phys. Lett. A}\ }\textbf {\bibinfo {volume}
  {156}},\ \bibinfo {pages} {441} (\bibinfo {year} {1991})}\BibitemShut
  {NoStop}%
\bibitem [{\citenamefont {Dahiya}, \citenamefont {John},\ and\ \citenamefont
  {Saxena}(1978)}]{Dahiya1978}%
  \BibitemOpen
  \bibfield  {author} {\bibinfo {author} {\bibfnamefont {R.~P.}\ \bibnamefont
  {Dahiya}}, \bibinfo {author} {\bibfnamefont {P.~I.}\ \bibnamefont {John}}, \
  and\ \bibinfo {author} {\bibfnamefont {Y.~C.}\ \bibnamefont {Saxena}},\
  }\href {\doibase 10.1016/0375-9601(78)90717-X} {\bibfield  {journal}
  {\bibinfo  {journal} {Phys. Lett. A}\ }\textbf {\bibinfo {volume} {65}},\
  \bibinfo {pages} {323} (\bibinfo {year} {1978})}\BibitemShut {NoStop}%
\bibitem [{\citenamefont {Nishida}(1984)}]{Nishida1984}%
  \BibitemOpen
  \bibfield  {author} {\bibinfo {author} {\bibfnamefont {Y.}~\bibnamefont
  {Nishida}},\ }\href {\doibase http://dx.doi.org/10.1063/1.864843} {\bibfield
  {journal} {\bibinfo  {journal} {Phys. Fluids}\ }\textbf {\bibinfo {volume}
  {27}},\ \bibinfo {pages} {2176} (\bibinfo {year} {1984})}\BibitemShut
  {NoStop}%
\bibitem [{\citenamefont {Imen}\ and\ \citenamefont {Kuehl}(1987)}]{Imen1987}%
  \BibitemOpen
  \bibfield  {author} {\bibinfo {author} {\bibfnamefont {K.}~\bibnamefont
  {Imen}}\ and\ \bibinfo {author} {\bibfnamefont {H.~H.}\ \bibnamefont
  {Kuehl}},\ }\href {\doibase 10.1063/1.866061} {\bibfield  {journal} {\bibinfo
   {journal} {Phys. Fluids}\ }\textbf {\bibinfo {volume} {30}},\ \bibinfo
  {pages} {73} (\bibinfo {year} {1987})}\BibitemShut {NoStop}%
\bibitem [{\citenamefont {Cooney}\ \emph {et~al.}(1991)\citenamefont {Cooney},
  \citenamefont {Gavin}, \citenamefont {Williams}, \citenamefont {Aossey},\
  and\ \citenamefont {Lonngren}}]{Cooney1991}%
  \BibitemOpen
  \bibfield  {author} {\bibinfo {author} {\bibfnamefont {J.~L.}\ \bibnamefont
  {Cooney}}, \bibinfo {author} {\bibfnamefont {M.~T.}\ \bibnamefont {Gavin}},
  \bibinfo {author} {\bibfnamefont {J.~E.}\ \bibnamefont {Williams}}, \bibinfo
  {author} {\bibfnamefont {D.~W.}\ \bibnamefont {Aossey}}, \ and\ \bibinfo
  {author} {\bibfnamefont {K.~E.}\ \bibnamefont {Lonngren}},\ }\href {\doibase
  10.1063/1.859759} {\bibfield  {journal} {\bibinfo  {journal} {Phys. Fluids
  B}\ }\textbf {\bibinfo {volume} {3}},\ \bibinfo {pages} {3277} (\bibinfo
  {year} {1991})}\BibitemShut {NoStop}%
\bibitem [{\citenamefont {Ohsawa}(1987)}]{Ohsawa1987}%
  \BibitemOpen
  \bibfield  {author} {\bibinfo {author} {\bibfnamefont {Y.}~\bibnamefont
  {Ohsawa}},\ }\href {\doibase 10.1143/JPSJ.56.433} {\bibfield  {journal}
  {\bibinfo  {journal} {J. Phys. Soc. Jpn.}\ }\textbf {\bibinfo {volume}
  {56}},\ \bibinfo {pages} {433} (\bibinfo {year} {1987})}\BibitemShut
  {NoStop}%
\bibitem [{\citenamefont {Toida}, \citenamefont {Dogen},\ and\ \citenamefont
  {Ohsawa}(1999)}]{Toida1999a}%
  \BibitemOpen
  \bibfield  {author} {\bibinfo {author} {\bibfnamefont {M.}~\bibnamefont
  {Toida}}, \bibinfo {author} {\bibfnamefont {D.}~\bibnamefont {Dogen}}, \ and\
  \bibinfo {author} {\bibfnamefont {Y.}~\bibnamefont {Ohsawa}},\ }\href
  {http://www.jspf.or.jp/JPFRS/PDF/Vol2/jpfrs1999_02-463.pdf} {\bibfield
  {journal} {\bibinfo  {journal} {Journal of Plasma and Fusion Research
  SERIES}\ }\textbf {\bibinfo {volume} {2}},\ \bibinfo {pages} {463} (\bibinfo
  {year} {1999})}\BibitemShut {NoStop}%
\bibitem [{\citenamefont {Heald}\ and\ \citenamefont
  {Wharton}(1965)}]{Heald1965}%
  \BibitemOpen
  \bibfield  {author} {\bibinfo {author} {\bibfnamefont {M.~A.}\ \bibnamefont
  {Heald}}\ and\ \bibinfo {author} {\bibfnamefont {C.~B.}\ \bibnamefont
  {Wharton}},\ }\href@noop {} {\emph {\bibinfo {title} {Plasma diagnostics with
  microwaves}}},\ \bibinfo {edition} {1st}\ ed.,\ Wiley series in plasma
  physics\ (\bibinfo  {publisher} {Wiley},\ \bibinfo {year} {1965})\BibitemShut
  {NoStop}%
\bibitem [{\citenamefont {Irie}\ and\ \citenamefont {Ohsawa}(2003)}]{Irie2003}%
  \BibitemOpen
  \bibfield  {author} {\bibinfo {author} {\bibfnamefont {S.}~\bibnamefont
  {Irie}}\ and\ \bibinfo {author} {\bibfnamefont {Y.}~\bibnamefont {Ohsawa}},\
  }\href {\doibase 10.1063/1.1568947} {\bibfield  {journal} {\bibinfo
  {journal} {Phys. Plasmas}\ }\textbf {\bibinfo {volume} {10}},\ \bibinfo
  {pages} {1253} (\bibinfo {year} {2003})}\BibitemShut {NoStop}%
\bibitem [{\citenamefont {Arber}\ \emph {et~al.}(2015)\citenamefont {Arber},
  \citenamefont {Bennett}, \citenamefont {Brady}, \citenamefont
  {Lawrence-Douglas}, \citenamefont {Ramsay}, \citenamefont {Sircombe},
  \citenamefont {Gillies}, \citenamefont {Evans}, \citenamefont {Schmitz},
  \citenamefont {Bell},\ and\ \citenamefont {Ridgers}}]{Arber2015}%
  \BibitemOpen
  \bibfield  {author} {\bibinfo {author} {\bibfnamefont {T.~D.}\ \bibnamefont
  {Arber}}, \bibinfo {author} {\bibfnamefont {K.}~\bibnamefont {Bennett}},
  \bibinfo {author} {\bibfnamefont {C.~S.}\ \bibnamefont {Brady}}, \bibinfo
  {author} {\bibfnamefont {A.}~\bibnamefont {Lawrence-Douglas}}, \bibinfo
  {author} {\bibfnamefont {M.~G.}\ \bibnamefont {Ramsay}}, \bibinfo {author}
  {\bibfnamefont {N.~J.}\ \bibnamefont {Sircombe}}, \bibinfo {author}
  {\bibfnamefont {P.}~\bibnamefont {Gillies}}, \bibinfo {author} {\bibfnamefont
  {R.~G.}\ \bibnamefont {Evans}}, \bibinfo {author} {\bibfnamefont
  {H.}~\bibnamefont {Schmitz}}, \bibinfo {author} {\bibfnamefont {A.~R.}\
  \bibnamefont {Bell}}, \ and\ \bibinfo {author} {\bibfnamefont {C.~P.}\
  \bibnamefont {Ridgers}},\ }\href {\doibase 10.1088/0741-3335/57/11/113001}
  {\bibfield  {journal} {\bibinfo  {journal} {Plasma Phys. Controlled Fusion}\
  }\textbf {\bibinfo {volume} {57}},\ \bibinfo {pages} {113001} (\bibinfo
  {year} {2015})}\BibitemShut {NoStop}%
\bibitem [{\citenamefont {Spitzer}(1962)}]{Spitzer1962}%
  \BibitemOpen
  \bibfield  {author} {\bibinfo {author} {\bibfnamefont {L.}~\bibnamefont
  {Spitzer}},\ }\href@noop {} {\emph {\bibinfo {title} {Physics of fully
  ionized gases}}}\ (\bibinfo  {publisher} {Interscience Publishers},\ \bibinfo
  {year} {1962})\BibitemShut {NoStop}%
\bibitem [{\citenamefont {Gueroult}\ and\ \citenamefont
  {Fisch}(2016)}]{Gueroult2016}%
  \BibitemOpen
  \bibfield  {author} {\bibinfo {author} {\bibfnamefont {R.}~\bibnamefont
  {Gueroult}}\ and\ \bibinfo {author} {\bibfnamefont {N.~J.}\ \bibnamefont
  {Fisch}},\ }\href {\doibase 10.1063/1.4943877} {\bibfield  {journal}
  {\bibinfo  {journal} {Phys. Plasmas}\ }\textbf {\bibinfo {volume} {23}},\
  \bibinfo {pages} {032113} (\bibinfo {year} {2016})}\BibitemShut {NoStop}%
\bibitem [{\citenamefont {Ohsawa}(2017{\natexlab{b}})}]{Ohsawa2017}%
  \BibitemOpen
  \bibfield  {author} {\bibinfo {author} {\bibfnamefont {Y.}~\bibnamefont
  {Ohsawa}},\ }\href {\doibase 10.1017/S0022377816001240} {\bibfield  {journal}
  {\bibinfo  {journal} {J. Plasma Phys.}\ }\textbf {\bibinfo {volume} {83}},\
  \bibinfo {pages} {595830101} (\bibinfo {year}
  {2017}{\natexlab{b}})}\BibitemShut {NoStop}%
\bibitem [{\citenamefont {Dudkin}\ \emph {et~al.}(1994)\citenamefont {Dudkin},
  \citenamefont {Nechaev}, \citenamefont {Peshkov}, \citenamefont {Ryzhkov},
  \citenamefont {Lukanin},\ and\ \citenamefont {Shlapakovskii}}]{Dudkin1994}%
  \BibitemOpen
  \bibfield  {author} {\bibinfo {author} {\bibfnamefont {G.~N.}\ \bibnamefont
  {Dudkin}}, \bibinfo {author} {\bibfnamefont {B.~A.}\ \bibnamefont {Nechaev}},
  \bibinfo {author} {\bibfnamefont {A.~V.}\ \bibnamefont {Peshkov}}, \bibinfo
  {author} {\bibfnamefont {V.~A.}\ \bibnamefont {Ryzhkov}}, \bibinfo {author}
  {\bibfnamefont {A.~A.}\ \bibnamefont {Lukanin}}, \ and\ \bibinfo {author}
  {\bibfnamefont {A.}~\bibnamefont {Shlapakovskii}},\ }\href@noop {} {\bibfield
   {journal} {\bibinfo  {journal} {JETP}\ }\textbf {\bibinfo {volume} {78}},\
  \bibinfo {pages} {865} (\bibinfo {year} {1994})}\BibitemShut {NoStop}%
\bibitem [{\citenamefont {Hietala}\ \emph {et~al.}(2011)\citenamefont
  {Hietala}, \citenamefont {Agueda}, \citenamefont {Andr{\'{e}}eov{\'{a}}},
  \citenamefont {Vainio}, \citenamefont {Nylund}, \citenamefont {Kilpua},\ and\
  \citenamefont {Koskinen}}]{Hietala2011}%
  \BibitemOpen
  \bibfield  {author} {\bibinfo {author} {\bibfnamefont {H.}~\bibnamefont
  {Hietala}}, \bibinfo {author} {\bibfnamefont {N.}~\bibnamefont {Agueda}},
  \bibinfo {author} {\bibfnamefont {K.}~\bibnamefont {Andr{\'{e}}eov{\'{a}}}},
  \bibinfo {author} {\bibfnamefont {R.}~\bibnamefont {Vainio}}, \bibinfo
  {author} {\bibfnamefont {S.}~\bibnamefont {Nylund}}, \bibinfo {author}
  {\bibfnamefont {E.~K.~J.}\ \bibnamefont {Kilpua}}, \ and\ \bibinfo {author}
  {\bibfnamefont {H.~E.~J.}\ \bibnamefont {Koskinen}},\ }\href {\doibase
  10.1029/2011ja016669} {\bibfield  {journal} {\bibinfo  {journal} {J. Geophys.
  Res.}\ }\textbf {\bibinfo {volume} {116}},\ \bibinfo {pages} {A10105}
  (\bibinfo {year} {2011})}\BibitemShut {NoStop}%
\bibitem [{\citenamefont {Dawson}(1964)}]{Dawson1964}%
  \BibitemOpen
  \bibfield  {author} {\bibinfo {author} {\bibfnamefont {J.~M.}\ \bibnamefont
  {Dawson}},\ }\href {\doibase 10.1063/1.1711214} {\bibfield  {journal}
  {\bibinfo  {journal} {Phys. Fluids}\ }\textbf {\bibinfo {volume} {7}},\
  \bibinfo {pages} {419} (\bibinfo {year} {1964})}\BibitemShut {NoStop}%
\bibitem [{\citenamefont {Okuda}\ and\ \citenamefont
  {Birdsall}(1970)}]{Okuda1970}%
  \BibitemOpen
  \bibfield  {author} {\bibinfo {author} {\bibfnamefont {H.}~\bibnamefont
  {Okuda}}\ and\ \bibinfo {author} {\bibfnamefont {C.~K.}\ \bibnamefont
  {Birdsall}},\ }\href {\doibase 10.1063/1.1693210} {\bibfield  {journal}
  {\bibinfo  {journal} {Phys. Fluids}\ }\textbf {\bibinfo {volume} {13}},\
  \bibinfo {pages} {2123} (\bibinfo {year} {1970})}\BibitemShut {NoStop}%
\bibitem [{\citenamefont {Montgomery}\ and\ \citenamefont
  {Nielson}(1970)}]{Montgomery1970}%
  \BibitemOpen
  \bibfield  {author} {\bibinfo {author} {\bibfnamefont {D.}~\bibnamefont
  {Montgomery}}\ and\ \bibinfo {author} {\bibfnamefont {C.~W.}\ \bibnamefont
  {Nielson}},\ }\href {\doibase 10.1063/1.1693081} {\bibfield  {journal}
  {\bibinfo  {journal} {Phys. Fluids}\ }\textbf {\bibinfo {volume} {13}},\
  \bibinfo {pages} {1405} (\bibinfo {year} {1970})}\BibitemShut {NoStop}%
\bibitem [{\citenamefont {Birdsall}\ and\ \citenamefont
  {Langdon}(1985)}]{Birdsall1985}%
  \BibitemOpen
  \bibfield  {author} {\bibinfo {author} {\bibfnamefont {C.~K.}\ \bibnamefont
  {Birdsall}}\ and\ \bibinfo {author} {\bibfnamefont {A.~B.}\ \bibnamefont
  {Langdon}},\ }\href@noop {} {\emph {\bibinfo {title} {Plasma Physics via
  Computer Simulation}}},\ edited by\ \bibinfo {editor} {\bibfnamefont {M.-H.}\
  \bibnamefont {Book}}\ (\bibinfo  {publisher} {McGraw-Hill},\ \bibinfo {year}
  {1985})\BibitemShut {NoStop}%
\end{thebibliography}

%

\appendix

\section{Korteweg-de-Vries (KdV) equation for a magnetosonic wave in a single ion species cold plasma}
\label{Sec:KdV}

The set of equations considered here is made of the continuity equation for electrons and ions, the momentum equation along $x$ and $y$ for electrons and ions, as well as Faraday's and Ampere's equations. The background magnetic field is $\mathbf{B} = \mathrm{B_0} \mathbf{\hat{z}}$ and the unperturbed plasma density is $\mathrm{n_0}$. The perturbation propagates along $\mathbf{\hat{x}}$. Under the assumptions $\partial/\partial \mathrm{y} = \partial/\partial \mathrm{z} = 0$ , it writes

\begin{subequations}
\begin{gather}
\frac{\partial \mathrm{n_e}}{\partial \mathrm{t}} + \frac{\partial \mathrm{n_e v_{ex}}}{\partial \mathrm{x}}  = 0, \\
\frac{\mathrm{\partial n_i}}{\partial \mathrm{t}} + \frac{\partial \mathrm{n_i v_{ix}}}{\partial \mathrm{x}}  = 0, \\
m_e\left(\frac{\partial }{\partial \mathrm{t}} + \mathrm{v_{ex}}\frac{\partial }{\partial \mathrm{x}}\right)\mathrm{v_{ex}} = -e (\mathrm{E_x + v_{ey} B_z}),\\
m_i\left(\frac{\partial }{\partial \mathrm{t}} + \mathrm{v_{ix}}\frac{\partial }{\partial \mathrm{x}}\right)\mathrm{v_{ix}} = e (\mathrm{E_x + v_{ey} B_z}),\\
m_e\left(\frac{\partial }{\partial \mathrm{t}} + \mathrm{v_{ex}}\frac{\partial }{\partial \mathrm{x}}\right)\mathrm{v_{ey}} = -e (\mathrm{E_y - v_{ex} B_z}),\\
m_i\left(\frac{\partial }{\partial \mathrm{t}} + \mathrm{v_{ix}}\frac{\partial }{\partial \mathrm{x}}\right)\mathrm{v_{iy}} = e (\mathrm{E_y - v_{ex} B_z}),\\
\frac{\partial \mathrm{B_z}}{\partial \mathrm{t}} = -\frac{\partial \mathrm{E_y}}{\partial \mathrm{x}},\\
\frac{\partial \mathrm{B_z}}{\partial \mathrm{x}} = -\mu_0 e (\mathrm{n_i v_{iy} - n_e v_{ey}}),
\end{gather}
\label{Eq:Set}
\end{subequations}
with $m_e$ and $m_i$ the electron and ion mass respectively, and $e$ the elementary charge. Here we introduced the normalized variables 
\begin{subequations}
\begin{gather}
x = \nicefrac{\mathrm{x}}{\frac{c}{\omega_{pe}}},\\
t = \nicefrac{\mathrm{t}}{\frac{c}{{v_A} \omega_{pe}}},\\
v_{i\alpha} = \nicefrac{\mathrm{v_{i\alpha}}}{v_A}  ,\quad v_{e\alpha} = \nicefrac{\mathrm{v_{e\alpha}}}{v_A}  \\
n = \nicefrac{\mathrm{n}}{\mathrm{n_0}},\\
B_z = \nicefrac{\mathrm{B_z}}{\mathrm{B_0}},\\
E_{\alpha} = \nicefrac{\mathrm{E_{\alpha}}}{v_A \mathrm{B_0}} ,
\end{gather}
\label{Eq:normalization}
\end{subequations}
where $\alpha$ designates $\mathrm{x}$ or $\mathrm{y}$, ${v_A} = \Omega_i/\omega_{pi} c$ is the Alfv{\'e}n velocity with $\Omega_i = e \mathrm{B_0}/m_i$ the ion cyclotron frequency, $\omega_{pi} = [\mathrm{n_0} e^2/(m_i \varepsilon_0)]^{1/2}$ the ion plasma frequency and $c$ the speed of light, and $\omega_{pe} = [\mathrm{n_0 }e^2/(m_e \varepsilon_0)]^{1/2}$ is the electron plasma frequency. Eqs.~(\ref{Eq:Set}) then reads
\begin{subequations}
\begin{gather}
\frac{\partial n_e}{\partial t} + \frac{\partial n_e v_{ex}}{\partial x}  = 0, \\
\frac{\partial n_i}{\partial t} + \frac{\partial n_i v_{ix}}{\partial x}  = 0, \\
\left(\frac{\partial }{\partial t} + v_{ex}\frac{\partial }{\partial x}\right)v_{ex} = -\eta^{-1} (E_x + v_{ey} B_z),\\
\left(\frac{\partial }{\partial t} + v_{ix}\frac{\partial }{\partial x}\right)v_{ix} = \eta (E_x + v_{ey} B_z),\\
\left(\frac{\partial }{\partial t} + v_{ex}\frac{\partial }{\partial x}\right)v_{ey} = -\eta^{-1} (E_y - v_{ex} B_z),\\
\left(\frac{\partial }{\partial t} + v_{ix}\frac{\partial }{\partial x}\right)v_{iy} = \eta (E_y - v_{ex} B_z),\\
\frac{\partial B_z}{\partial t} = -\frac{\partial E_y}{\partial x},\\
\frac{\partial B_z}{\partial x} = -\eta (n_i v_{iy} - n_e v_{ey}),
\end{gather}
\label{Eq:SetNormalized}
\end{subequations}
with $\eta = (m_e/m_i)^{1/2}$ the square root of the mass ratio. We now introduce the stretched coordinates
\begin{subequations}
\begin{gather}
\xi = \epsilon^{1/2} (x-t) \\
\tau = \epsilon^{3/2} t, 
\end{gather}
\label{Eq:VariableChange}
\end{subequations}
so that
\begin{alignat}{2}
 & \frac{\partial (\cdot)}{\partial x}    \quad&\rightarrow \quad&\epsilon^{1/2}\frac{\partial (\cdot)}{\partial \xi}, \\	
& \frac{\partial (\cdot)}{\partial t}    \quad&\rightarrow \quad&-\epsilon^{1/2}\frac{\partial (\cdot)}{\partial \xi}+\epsilon^{3/2}\frac{\partial (\cdot)}{\partial \tau},
\end{alignat}
and expand the plasma variables as
\begin{subequations}
\begin{gather}
B_z = 1 + \epsilon B_{z_1} + \epsilon^2 B_{z_2} + \cdots, \\
n_i = 1 + \epsilon n_{i_1} + \epsilon^2 n_{i_2} + \cdots, \\
n_e = 1 + \epsilon n_{e_1} + \epsilon^2 n_{e_2} + \cdots, \\
v_{ex} = \epsilon v_{ex_1} + \epsilon^2 v_{ex_2} + \cdots, \\
v_{ix} = \epsilon v_{ix_1} + \epsilon^2 v_{ix_2} + \cdots, \\
E_y = \epsilon E_{y_1} + \epsilon^2 E_{y_2} + \cdots, \\
E_x = \eta^{-1}(\epsilon^{3/2} E_{x_1} + \epsilon^{5/2} E_{x_2} + \cdots), \\
v_{ey} = \eta^{-1}(\epsilon^{3/2} v_{ey_1} + \epsilon^{5/2} v_{ey_2} + \cdots), \\
v_{iy} = \eta^{-1}(\epsilon^{3/2} v_{iy_1} + \epsilon^{5/2} v_{iy_2} + \cdots). 
\end{gather}
\label{Eq:Expansion}
\end{subequations}
Plugging Eqs.~(\ref{Eq:Expansion}) into Eqs.~(\ref{Eq:SetNormalized}) yields
\begin{subequations}
\begin{multline}
\epsilon^{3/2} \left[ -\frac{\partial n_{e_1}}{\partial \xi}+\frac{\partial v_{ex_1}}{\partial \xi}\right] \\+ \epsilon^{5/2} \left[ -\frac{\partial n_{e_2}}{\partial \xi}+\frac{\partial n_{e_1}}{\partial \tau}+\frac{\partial v_{ex_2}}{\partial \xi}+\frac{\partial n_{e_1}v_{ex_1}}{\partial \xi}\right] \\+ \cdots = 0,
\end{multline}
\begin{multline}
\epsilon^{3/2} \left[ -\frac{\partial n_{i_1}}{\partial \xi}+\frac{\partial v_{ix_1}}{\partial \xi}\right] \\+ \epsilon^{5/2} \left[ -\frac{\partial n_{i_2}}{\partial \xi}+\frac{\partial n_{i_1}}{\partial \tau}+\frac{\partial v_{ix_2}}{\partial \xi}+\frac{\partial n_{i_1}v_{ix_1}}{\partial \xi}\right] \\+ \cdots = 0,
\end{multline}
\begin{multline}
\eta^{-2}\epsilon^{3/2}\left[E_{x_1}+v_{ey_1}\right] + \eta^{-2}\epsilon^{5/2}\left[E_{x_2}+v_{ey_2} + v_{ey_1} B_{z_1}\right] \\+  \cdots = 0,\label{Eq:electron_momentum_x}
\end{multline}
\begin{multline}
\epsilon^{3/2}\left[\frac{\partial v_{ix_1}}{\partial \xi}+ E_{x_1}+v_{iy_1}\right] \\+ \epsilon^{5/2}\left[\frac{\partial v_{ix_2}}{\partial \xi}-\frac{\partial v_{ix_1}}{\partial \tau}-v_{ix_1}\frac{\partial v_{ix_1}}{\partial \xi}+ E_{x_2}+v_{iy_2}+ v_{iy_1}B_{z_1}\right] \\
+ \cdots = 0,
\label{Eq:ion_momentum_x}
\end{multline}
\begin{multline}
\eta^{-1}\epsilon\left[E_{y_1}-v_{ex_1}\right] + \\\eta^{-1}\epsilon^{2}\left[-\frac{\partial v_{ey_1}}{\partial \xi} +E_{y_2}-v_{ex_2} - v_{ex_1} B_{z_1}\right] +\cdots = 0,\label{Eq:electron_momentum_y}
\end{multline}
\begin{multline}
\eta^{-1}\epsilon^{2}\frac{\partial v_{iy_1}}{\partial \xi}+\eta^{-1}\epsilon^{3}\left[\frac{\partial v_{iy_2}}{\partial \xi} -\frac{\partial v_{iy_1}}{\partial \tau} -v_{ix_1}\frac{\partial v_{iy_1}}{\partial \xi}\right]\\+\eta\epsilon\left[E_{y_1}-v_{ix_1}\right]
+\eta\epsilon^2\left[E_{y_2}-v_{ix_2}-v_{ix_1}B_{z_1}\right] \\+ \cdots = 0,
\end{multline}
\begin{multline}
\epsilon^{3/2}\left[-\frac{\partial B_{z_1}}{\partial \xi}+\frac{\partial E_{y_1}}{\partial \xi}\right]+\epsilon^{5/2}\left[-\frac{\partial B_{z_2}}{\partial \xi}+\frac{\partial B_{z_1}}{\partial \tau}+\frac{\partial E_{y_2}}{\partial \xi}\right]\\+ \cdots = 0,\label{Eq:Faraday}
\end{multline}
\begin{multline}
\epsilon^{3/2}\left[\frac{\partial B_{z_1}}{\partial \xi}+v_{ix_1}-v_{ex_1}\right]\\+\epsilon^{5/2}\left[\frac{\partial B_{z_2}}{\partial \xi}+n_{i_1}v_{ix_1}-n_{e_1}v_{ex_1}+v_{ix_2}-v_{ex_2}\right]\\+ \cdots = 0.
\label{Eq:Ampere}
\end{multline}
\label{Eq:LowestOrder}
\end{subequations}

From lowest order terms in Eqs.~(\ref{Eq:LowestOrder}), one gets
\begin{subequations}
\begin{gather}
n_{e_1} = v_{ex_1} = E_{y_1}  = B_{z_1}\\
n_{i_1} = v_{ix_1}.
\end{gather}
\label{Eq:res_lowest_order}
\end{subequations}
Now, the choice of a given plasma composition determines $\eta = (m_e/m_i)^{1/2}$. For an electron/proton plasma, $\eta\sim1/43$. For a soliton amplitude such that $\eta \ll \epsilon\ll 1$,  lowest order terms in Eqs.~(\ref{Eq:LowestOrder}) further gives
\begin{subequations}
\begin{gather}
v_{iy_1} = 0\\
v_{ix_1} = B_{z_1},\\
v_{ey_1} = -E_{x_1} = \partial B_{z_1}/\partial \xi. 
\end{gather}
\label{Eq:res_lowest_order2}
\end{subequations}

The $\epsilon$ and $\epsilon^2$ terms in Ampere's law expansion along the $x$ direction gives respectively $v_{ex_1} = v_{ix_1}$ and $v_{ex_2} = v_{ix_2}$. Plugging these results into the $\mathcal{O} (\epsilon^{5/2})$ term of Eq.~(\ref{Eq:ion_momentum_x}) yields
\begin{equation}
\frac{\partial B_{z_1}}{\partial \tau} -\frac{\partial v_{ex_2}}{\partial \xi} +\frac{1}{2}\frac{\partial ({B_{z_1}})^2}{\partial \xi} -E_{x_2} = 0
\label{Eq:intermediate}
\end{equation}
where use has been made of Eqs.~(\ref{Eq:res_lowest_order}). Eq.~(\ref{Eq:intermediate}), together with the $\mathcal{O} (\eta^{-2}\epsilon^{5/2})$ term in Eq.~(\ref{Eq:electron_momentum_x}), the $\mathcal{O} (\eta^{-1}\epsilon^{2})$ term in Eq.~(\ref{Eq:electron_momentum_y}), the $\mathcal{O} (\epsilon^{5/2})$ term in Eq.~(\ref{Eq:Faraday}) and the $\mathcal{O} (\epsilon^{5/2})$ term in Eq.~(\ref{Eq:Ampere}), are then used to eliminate second order coefficients $v_{ex_2}$, $v_{ey_2}$, $E_{x_2}$, $E_{y_2}$ and $B_{z_2}$ to  yield the evolution equation for $B_{z_1}$,
\begin{equation}
\frac{\partial B_{z_1}}{\partial \tau} + \frac{3}{2}B_{z_1}\frac{\partial B_{z_1}}{\partial \xi} + \frac{1}{2}\frac{\partial^3 B_{z_1}}{\partial \xi^3} = 0.
\label{Eq:KdV}
\end{equation}
Eq.~(\ref{Eq:KdV}) is the Korteweg-de Vries (KdV) equation for a perpendicular magnetosonic solitary wave in a single ion species cold plasma. The general solution to Eq.~(\ref{Eq:KdV}) is
\begin{equation}
B_{z_1} = a \sech^2\left[\frac{\sqrt{a}}{2}\left(\xi-\frac{a}{2}\tau\right)\right],
\label{Eq:KdVSolution}
\end{equation}
with $a \in \mathbf{I\!R}_{>0}$. Eq.~(\ref{Eq:KdVSolution}) shows that the only soliton solutions for perpendicular propagation are compressive solitons. Rarefaction solitons only exist for oblique propagation when $\cos(\theta)>\eta$~\cite{Ohsawa2014}, with $\theta$ the angle between the magnetic field and the pulse propagation direction. Returning to the dimensionless variables $x$ and $t$ defined in Eqs.~(\ref{Eq:VariableChange}), Eq.~(\ref{Eq:KdVSolution}) gives
\begin{equation}
B_{z}(x,t) = 1+\epsilon\sech^2\left(\frac{\sqrt{\epsilon}}{2}  \left[x-t (1+\epsilon/2) \right] \right) + \mathcal{O}(\epsilon^2).
\label{Eq:BzSoliton}
\end{equation}
From Eqs.~(\ref{Eq:Expansion}), (\ref{Eq:res_lowest_order}) and (\ref{Eq:res_lowest_order2}), one similarly obtains the wave electric field components $E_x$ and $E_y$,
\begin{subequations}
\begin{multline}
E_x = \eta^{-1}\left[\epsilon^{3/2}\sech^2\left(\frac{\sqrt{\epsilon}}{2}  \left[x-t (1+\epsilon/2) \right] \right)\right.\\\left.\times\tanh\left(\frac{\sqrt{\epsilon}}{2}  \left[x-t (1+\epsilon/2) \right] \right)+ \mathcal{O}(\epsilon^{5/2})\right]
\label{Eq:ExSoliton}
\end{multline}
\begin{equation}
E_y = \epsilon\sech^2\left(\frac{\sqrt{\epsilon}}{2}  \left[x-t (1+\epsilon/2) \right] \right)+ \mathcal{O}(\epsilon^2),
\label{Eq:EySoliton}
\end{equation}
\label{Eq:fields}
\end{subequations}
and the density and fluid velocities,
\begin{subequations}
\begin{gather}
n = n_e = n_i = 1+\epsilon\sech^2\left(\frac{\sqrt{\epsilon}}{2}  \left[x-t (1+\epsilon/2) \right] \right)+ \mathcal{O}(\epsilon^2),\\
v_x = v_{ex} = v_{ix} = \epsilon\sech^2\left(\frac{\sqrt{\epsilon}}{2}  \left[x-t (1+\epsilon/2) \right] \right)+ \mathcal{O}(\epsilon^2),
\label{Eq:VxeSoliton}
\end{gather}
\begin{multline}
v_{ey} = -\eta^{-1}\left[\epsilon^{3/2}\sech^2\left(\frac{\sqrt{\epsilon}}{2}  \left[x-t (1+\epsilon/2) \right] \right)\right.\\\left.\times\tanh\left(\frac{\sqrt{\epsilon}}{2}  \left[x-t (1+\epsilon/2) \right] \right)+ \mathcal{O}(\epsilon^{5/2})\right].
\label{Eq:VyeSoliton}
\end{multline}
\label{Eq:fields2}
\end{subequations}
In dimensional units $\mathrm{x}$ and $\mathrm{t}$, the magnetic field reads
\begin{multline}
\mathrm{B_{z}}(\mathrm{x},\mathrm{t}) = \mathrm{B_0}\left[1+\epsilon\sech^2\left(\frac{\omega_{pe}}{2c}\sqrt{\epsilon}  \left[\mathrm{x}-{v_A} \mathrm{t} (1+\epsilon/2) \right] \right)\right.\\\left.+ \mathcal{O}(\epsilon^2)\vphantom{\frac12}\right].
\end{multline}

\section{Soliton energy}
\label{Sec:Energy}

The field and particle energy contents associated with the pulse are defined as
\begin{equation}
\varepsilon_\mathcal{F}(t) = \int_{-L_p}^{L_p} \left(\frac{\epsilon_0}{2} \left[E_x^2+E_y^2\right]+\frac{1}{2\mu_0}\left[B_z-1\right]^2\right)dx,
\label{Eq:SolitonFieldEnergy}
\end{equation}
and 
\begin{equation}
\varepsilon_\mathcal{P} (t)= {v_A}\int_{-L_p/2}^{L_p/2} \frac{n m_i}{2}\left[(1+\eta^2){v_x}^2+\left({v_{iy}}^2+\eta^2{v_{ey}}^2\right)\right]dx.
\label{Eq:SolitonParticleEnergy}
\end{equation}
At $t=0$, these quantities can be estimated from the first order expansion of the solution of the KdV equation given in Appendix~\ref{Sec:KdV}, and read
\begin{equation}
\varepsilon_\mathcal{F} = \frac{4}{3}\frac{{B_0}^2(B_m-1)^{3/2}}{\mu_0}\frac{c}{\omega_{pe}}\left[1+\frac{1}{5}\frac{(B_m-1)}{\eta^2}\frac{{v_A}^2}{c^2}+\frac{{v_A}^2}{c^2}\right]
\end{equation}
and
\begin{multline}
\varepsilon_\mathcal{P} = \frac{4}{3}\frac{{B_0}^2(B_m-1)^{3/2}}{\mu_0}\frac{c}{\omega_{pe}}\\\times\left[1+(B_m-1)+\frac{4}{35}(B_m-1)^2\right.\\\left.+\eta^2+\frac{4}{5}(B_m-1)\eta^2\right],
\end{multline} 
where we used the results 
\begin{subequations}
\begin{gather}
\int_{-\infty}^{\infty}\sech^4(x)dx = 4/3,\\
\int_{-\infty}^{\infty}\sech^4(x)\tanh^2(x)dx = 4/15,\\
\int_{-\infty}^{\infty}\sech^6(x)dx = 16/15,\\
\int_{-\infty}^{\infty}\sech^6(x)\tanh^2(x)dx = 16/105.
\end{gather}
\end{subequations}
To the lowest order in soliton amplitude $(B_m-1)$, the field and particles energy contents are equal, with 
\begin{equation}
{\varepsilon_\mathcal{F}}^0 = {\varepsilon_\mathcal{P}}^0 = \frac{4}{3}\frac{{B_0}^2(B_m-1)^{3/2}}{\mu_0}\frac{c}{\omega_{pe}}.
\label{Eq:Energy_content}
\end{equation}
We note that for the over-dense regime considered here ($\eta^{-1} {v_A}^2/c^2\ll1$), the particle energy content is $B_m$ times larger than the field energy content.

\section{Thermalization and grid-effects}
\label{Sec:Grid_Effects}

\begin{figure*}
\begin{center}
\subfigure[~$T_e/T_{e_0}$ over the entire domain]{\includegraphics[]{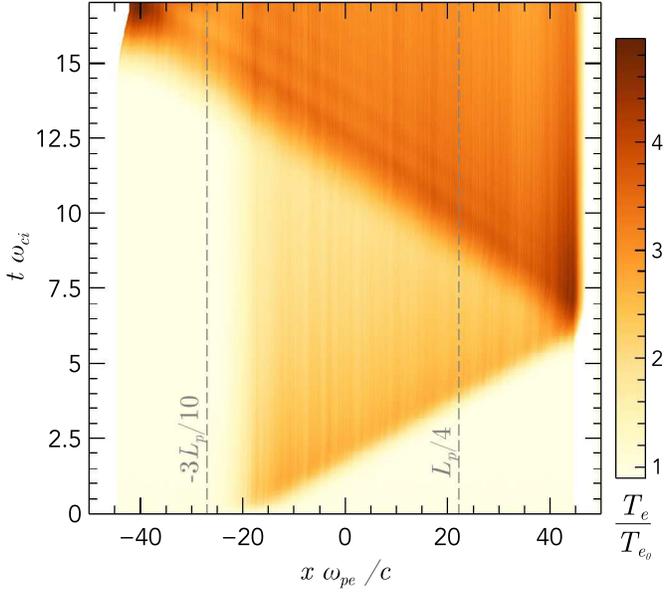}\label{Fig:HeatingMap}}\subfigure[~Electron and ion temperature at two positions]{\includegraphics[]{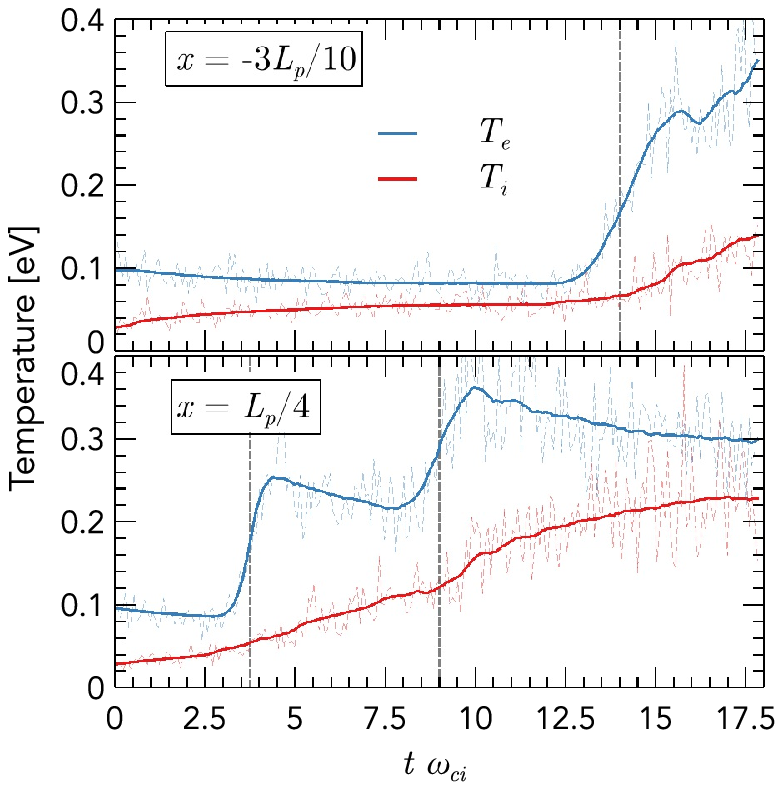}\label{Fig:TempEvol}}
\caption{Time evolution of the electron and ion temperature. Fig.~\subref{Fig:HeatingMap} shows the relative evolution over the entire simulation domain, whereas Fig.~\subref{Fig:TempEvol} displays the ion and electron temperature evolution at two different positions. $T_{e_0} = 0.1$~eV and $T_{e_0} = 0.03$~eV are the initial electron and ion temperature. The vertical dotted-grey lines in \subref{Fig:HeatingMap} denote the positions used in Fig.~\subref{Fig:TempEvol}.}
\label{Fig:Heating}
\end{center}
\end{figure*}

Fig.~\ref{Fig:HeatingMap} depicts the evolution of the relative electron temperature $T_e/T_{e_0}$, with $T_{e_0} = 0.1$~eV, over the entire simulation domain, while Fig.~\ref{Fig:TempEvol} shows the evolution of the electron and ion temperature at two specific positions indicated in dotted-grey in Fig.~\ref{Fig:HeatingMap}. Because of the ordering $\tau_r\leq{\tau_{ie}}^{\varepsilon}\leq L_p/(M_A V_A)$, these results can be analyzed in two steps: the modifications induced by the pulse on one hand, and the plasma evolution in between passages of the pulse on the other hand.

For each of the pulse passage ($t\omega_{ci}\sim3.5$ and $t\omega_{ci}\sim9$ for $x=L_p/4$ and $t\omega_{ci}\sim14$ for $x=-3L_p/10$), a step increase in $T_e$ is clearly visible in Fig.~\ref{Fig:TempEvol}. This increase is consistent with the energy deposition by the pulse discussed in Sec.~\ref{Sec:PIC}. Fig.~\ref{Fig:TempEvol} also confirms that the energy deposited by the pulse is essentially transferred to the electrons. This is particularly true for the first passage ($t\omega_{ci}\sim3.5$) when the pulse's shape is still very close to a magnetosonic soliton and the upstream plasma remains undisturbed.

In between passages of the pulse, Fig.~\ref{Fig:TempEvol} shows that electrons cool down on ions as expected from electron-ion ($e-i$) collisions in the regime where   ${\tau_{ie}}^{\varepsilon}\leq L_p/(M_A V_A)$. However, $e-i$ collisions are not included in this PIC model. Here, energy relaxation occurs as a result of grid effects. Indeed, finite size particles in PIC models are known to lead to spurious numerical thermalization~\cite{Dawson1964,Okuda1970}. The rate of this unphysical energy relaxation $\tau_N$ depends on $N_D$, the number of simulated particles in a Debye sphere, and $\tau_N\propto {N_D}^2$ has been verified for one-dimensional simulations~\cite{Montgomery1970}. Based on Fig.~\ref{Fig:TempEvol}, the characteristic time for numerical thermalization $\tau_N\sim{\omega_{ci}}^{-1}$, which is a few times larger than Spitzer's equipartition time $\tau_S$ given in Eq.~(\ref{Eq:Spitzer}). Grid effects therefore reproduce thermalization from $e-i$ collisions but underestimate the equipartition rate. 

Enforcing the physical equipartition rate may have an impact on the simulation results. Yet, although the initial pulse encounters plasma with electron and ion temperatures varying by up to $20\%$ as a result of numerical thermalization along its first pass towards the right boundary, it does not show significant changes in properties. In particular, the ion displacement computed for ions with initial positions $0.05\leq x_0/L\leq 0.35$, \emph{i.~e.} ions with varying temperature when reached by the pulse, shows little relative variation ($\sim 1\%$). This suggests that the plasma's distance from thermal equilibrium, and therefore the equipartition rate, does not play a dominant effect in this plasma regime, or at least not on the ion displacement prediction.  

Note that numerical thermalization is a separate effect from the numerical spurious heating observed with electrostatic particle-in-cell codes when the Debye length is not resolved~\cite{Birdsall1985}. Numerical thermalization occurs even if the grid size is smaller than the Debye length, as it is in the simulations presented here.

\end{document}